\documentclass[a4paper,10pt]{article}

\usepackage{latexsym}
\usepackage{amsmath}
\usepackage{amssymb}
\usepackage{amscd}

\newlength{\minitwocolumn}
\makeatletter
\@addtoreset{equation}{section}
\makeatother

\title{Vertex operator approach for 
form factors of \\ 
Belavin's $(\mathbb{Z}/n\mathbb{Z})$-symmetric model}

\author{Yas-Hiro Quano\thanks
{email: quanoy@suzuka-u.ac.jp}}

\date{\it Department of Clinical Engineering, 
Suzuka University of Medical Science \\
      \it Kishioka-cho, Suzuka 510-0293, Japan \\
2 December 2009 \\[1cm]
Dedicated to the memory of my parents}
\begin{document}

\maketitle
\begin{abstract}
Belavin's $(\mathbb{Z}/n\mathbb{Z})$-symmetric model is considered 
on the basis of bosonization of vertex operators 
in the $A^{(1)}_{n-1}$ model and vertex-face transformation. 
Free field representations of nonlocal tail operators are constructed for 
off diagonal matrix elements with respect to the ground state sectors. 
As a result, integral formulae for form factors of any local operators 
in the $(\mathbb{Z}/n\mathbb{Z})$-symmetric model can be obtained, 
in principle. 
\end{abstract}

\section{Introduction}

The present paper is a continuation of \cite{Bel-corr}, in which 
we derived the integral formulae for correlation functions of 
Belavin's $(\mathbb{Z}/n\mathbb{Z})$-symmetric model \cite{Bela,RT} 
on the basis of vertex operator approach \cite{JMbk}. 
Belavin's $(\mathbb{Z}/n\mathbb{Z})$-symmetric model 
is an $n$-state generalization of 
Baxter's eight-vertex model \cite{ESM}, which has 
$(\mathbb{Z}/2\mathbb{Z})$-symmetries. 
As for the eight-vertex model, the integral formulae for correlation functions 
were derived by Lashkevich and Pugai \cite{LaP}, and those for form factors 
were derived by Lashkevich \cite{La}. 

It was found in \cite{LaP} that 
the correlation functions of the eight-vertex model can be obtained by 
using the free field realization of the vertex operators in the 
eight-vertex SOS model \cite{LuP}, with insertion of the nonlocal opearator 
$\Lambda$, called `the tail operator'. 
The most essential part of \cite{LaP} was 
the construction of free field representations of $\Lambda$'s. 
Furthermore, those of the off-diagoal 
(with respect to the ground state sector) elements of $\Lambda$'s 
were constructed in \cite{La}, in order to obtain the form factor 
formulae of the eight-vertex model. 

There are some researches which generalize the study of \cite{LaP}. 
The vertex operator approach for higher spin generalization of the 
eight-vertex model was presented in \cite{KKW}. 
For higher rank generalization, the integral formulae 
for correlation functions of 
Belavin's $(\mathbb{Z}/n\mathbb{Z})$-symmetric model were presented 
in our previous paper \cite{Bel-corr}. 
The expression of the spontaneous polarization of 
$(\mathbb{Z}/n\mathbb{Z})$-symmetric model \cite{SPn} was also reproduced 
in \cite{Bel-corr}, on the basis of vertex operator approach. 
To the best of our knowledge, 
there has been no developed research of \cite{La} related to the form factor 
problem. The aim of the present paper is to give a higher rank generalization 
of the bosonization scheme in the eight-vertex model. 

The present paper is organized as follows. In section 2 
we review the basic definitions of 
$(\mathbb{Z}/n\mathbb{Z})$-symmetric model \cite{Bela}, 
the corresponding dual face model $A^{(1)}_{n-1}$-model \cite{JMO}, 
and the vertex-face correspondence. Some detail definitions 
of the models concerned are listed in Appendix A. 
In section 3 we introduce the 
type I and type II vertex operators of both 
$(\mathbb{Z}/n\mathbb{Z})$-symmetric model and the $A^{(1)}_{n-1}$-model, 
and also introduce the tail operators. Furthermore, we derive 
the commutation relations that those operators should satisfy. 
In order to obtain integral formulae for 
form factors of $(\mathbb{Z}/n\mathbb{Z})$-symmetric model 
we construct the free field representations of off-diagonal elements 
of the tail operators, by using those 
of the type I \cite{AJMP} and the type II \cite{FHSY,FKQ} 
vertex operators in the $A^{(1)}_{n-1}$-model in section 4. 
Useful operator product expansion (OPE) formulae and 
commutation relations for basic bosons 
are given in Appendix B. 
In section 5 we give some concluding remarks. Among these remarks, 
a brief proof of the commutation relations of 
the type I and the type II vertex operators in the $A^{(1)}_{n-1}$-model 
is given in Appendix C. 

\section{Basic definitions} 

The present section aims to formulate 
the problem, thereby fixing the notation. 

\subsection{Theta functions} 

Jacobi theta function with two pseudo-periods $1$ and 
$\tau$\,(${\rm Im}\,\tau >0$) are defined as follows: 
\begin{equation}
\vartheta\left[\begin{array}{c} a \\ b \end{array} \right]
(v;\tau ): =\displaystyle\sum_{m\in \mathbb{Z}} 
\exp \left\{ \pi \sqrt{-1}(m+a)~\left[ (m+a)\tau 
+2(v+b) \right] \right\}, \label{Rieth}
\end{equation}
for $a,b\in\mathbb{R}$. 
Let $n\in\mathbb{Z}_{\geqslant 2}$ and 
$r\in\mathbb{R}$ such that $r> n-1$, and also fix 
the parameter $x$ such that $0<x<1$. 
We will use the abbreviations, 
\begin{equation}
[v]=x^{\frac{v^2}{r}-v}\Theta_{x^{2r}}(x^{2v}), 
~~~~
[v]'=x^{\frac{v^2}{r-1}-v}\Theta_{x^{2r-2}}(x^{2v}), 
\end{equation}
where 
\begin{eqnarray*}
&&\Theta_{q}(z)=(z; q)_\infty 
(qz^{-1}; q)_\infty (q; q)_\infty =
\sum_{m\in\mathbb{Z}} q^{m(m-1)/2}(-z)^m, \\
&&(z; q_1 , \cdots , q_m )_\infty = 
\prod_{i_1 , \cdots , i_m \geqslant 0} 
(1-zq_1^{i_1} \cdots q_m^{i_m}). 
\end{eqnarray*}
Note that 
$$
\vartheta\left[\begin{array}{c} 1/2 \\ -1/2 \end{array} \right]
\left( \frac{v}{r}, \frac{\pi\sqrt{-1}}{\epsilon r} \right)
=\sqrt{\frac{\epsilon r}{\pi}}
\exp\,\left(-\frac{\epsilon r}{4}\right)[v], 
$$
where $x=e^{-\epsilon}$ ($\epsilon >0$). 

For later conveniences we also 
introduce the following symbols
\begin{eqnarray}
r_{j}(v)&=&z^{\frac{r-1}{r}\frac{n-j}{n}}
\frac{g_{j}(z^{-1})}{g_{j}(z)}, ~~~~
g_{j}(z)=
\frac{\{x^{2n+2r-j-1}z\}
\{x^{j+1}z\}}
{\{x^{2n-j+1}z\}\{x^{2r+j-1}z\}}, \label{eq:g-def} \\
r^*_{j}(v)&=&z^{\frac{r}{r-1}\frac{n-j}{n}}
\frac{g^*_{j}(z^{-1})}{g^*_{j}(z)}, ~~~~
g^*_{j}(z)=
\frac{\{x^{2n+2r-j-1}z\}'
\{x^{j-1}z\}'}
{\{x^{2n-j-1}z\}'\{x^{2r+j-1}z\}'}, \label{eq:g*-def} \\
\chi_j (v)&=&z^{-\frac{j(n-j)}{n}} \dfrac{\rho_j (z^{-1})}{
\rho_j (z)}, ~~~~ 
\rho_j (z)=\frac{(-x^{2j+1}z;x^2,x^{2n})_\infty
(-x^{2n-2j+1}z;x^2,x^{2n})_\infty
}{(-xz;x^2,x^{2n})_\infty
(-x^{2n+1}z;x^2,x^{2n})_\infty
} \label{eq:chi-def}
\end{eqnarray}
where $z=x^{2v}$, $1\leqslant j\leqslant n$ and 
\begin{equation}
\{z\}=(z;x^{2r},x^{2n})_\infty , ~~~~
\{z\}'=(z;x^{2r-2},x^{2n})_\infty . 
\label{eq:{z}}
\end{equation}
In particular we denote $\chi (v)=\chi_1 (v)$. 
These factors will appear in the commutation relations 
among the type I and type II vertex operators. 

The integral kernel for the type I and the type II 
vertex operators will be given as the products of 
the following elliptic functions 
\begin{eqnarray}
f(v,w)=\frac{[v+\frac{1}{2}-w]}{[v-\frac{1}{2}]},~~~~
h(v)=\frac{[v-1]}{[v+1]}, \label{eq:fh-def} \\
f^*(v,w)=\frac{[v-\frac{1}{2}+w]'}{[v+\frac{1}{2}]'},~~~
h^*(v)=\frac{[v+1]'}{[v-1]'}. \label{eq:f*h*-def}
\end{eqnarray}
In section 4 we use the following identities 
\begin{equation}
\sum_{\nu =0}^{n-1} \prod_{j=0\atop j\neq\nu}^{n-1} 
\dfrac{f(v_{j+1} -v_j, 1-p_{\nu} +p_j)}{[p_{\nu}-p_j]}=0, 
\label{eq:sum=0}
\end{equation}
and 
\begin{equation}
\sum_{\nu =0}^{n-1} \prod_{j=0\atop j\neq\nu}^{n-1} 
\dfrac{f^*(v_j-v_{j+1}, 1-p_{j}+p_\nu )}{[p_{\nu}-p_j]'}=0, 
\label{eq:sum=0^*}
\end{equation}
where $v_n =v+\frac{n}{2}$, and $\displaystyle\sum_{
j=0}^{n-1} p_j =0$. 
The former one (\ref{eq:sum=0}) was derived in \cite{AJMP} 
by applying the Liouville's second theorem to 
the following elliptic function 
$$
F(w)=\prod_{j=0}^{n-1} \dfrac{[v_{j+1}-v_j-\frac{1}{2}+w-p_j ]}
{[v_{j+1}-v_j-\frac{1}{2}][w-p_j ]}. 
$$
The latter one (\ref{eq:sum=0^*}) can be similarly proved. 

\subsection{$(\mathbb{Z}/n\mathbb{Z})$-symmetric model 
and its dual face model}

Let $V=\mathbb{C}^n$ and 
$\{ \varepsilon _\mu \}_{0 \leqslant \mu \leqslant n-1}$ be 
the standard orthonormal basis with the inner 
product $\langle \varepsilon _\mu , 
\varepsilon _\nu \rangle =\delta_{\mu \nu}$. 
Belavin's $(\mathbb{Z}/n\mathbb{Z})$-symmetric model \cite{Bela} is 
a vertex model on a two-dimensional square lattice ${\cal L}$ 
such that the state variables take the values of 
$(\mathbb{Z}/n\mathbb{Z})$-spin. The model is 
$(\mathbb{Z}/n\mathbb{Z})$-symmetric in a sense that the $R$-matrix satisfies 
the following conditions: 
\begin{equation}
\begin{array}{cl}
\mbox{({\romannumeral 1})} & R(v)^{ik}_{jl}=0, 
\mbox{~~unless $i+k=j+l$,~~mod $n$}, \\[2mm]
\mbox{({\romannumeral 2})} & R(v)^{i+p k+p}_{j+p l+p}=
R(v)^{ik}_{jl}, 
\mbox{~$\forall i,j,k,l,p\in \mathbb{Z}/n\mathbb{Z}$}.
\end{array} \label{Znsym}
\end{equation}
The $R$-matrix satisfies the Yang-Baxter equation (YBE) 
\begin{eqnarray}
R_{12}(v_1 -v_2 )
R_{13}(v_1 -v_3 )
R_{23}(v_2 -v_3 )=
R_{23}(v_2 -v_3 )
R_{13}(v_1 -v_3 )
R_{12}(v_1 -v_2 ), \label{YBE}
\end{eqnarray}
where $R_{ij}(v)$ denotes the matrix on $V^{\otimes 3}$, 
which acts as $R(v)$ on the $i$-th and $j$-th components and 
as identity on the other one. As for the elliptic parametrization 
of $R$-matrix, see Appendix A. 

The dual face model of $(\mathbb{Z}/n\mathbb{Z})$-symmetric model 
is called $A^{(1)}_{n-1}$-model. This is a face model 
on a two-dimensional square lattice ${\cal L^*}$, the dual lattice 
of ${\cal L}$, 
such that the state variables take the values of 
the dual space of Cartan subalgebra ${\mathfrak h}^*$ of $A^{(1)}_{n-1}$: 
\begin{equation}
{\mathfrak h}^*=\bigoplus_{\mu =0}^{n-1} 
\mathbb{C} \omega_\mu , 
\label{eq:wt-space}
\end{equation}
where 
$$
\omega_\mu :=\sum_{\nu =0}^{\mu -1} \bar{\varepsilon}_\nu , 
~~~~ 
\bar{\varepsilon}_\mu =
\varepsilon _\mu -\frac{1}{n}\sum_{\mu =0}^{n-1} 
\varepsilon _\mu . 
$$
The weight lattice $P$ and the root lattice $Q$ of 
$A^{(1)}_{n-1}$ are usually defined. 
See Appendix A. 

An ordered pair $(a,b) \in {\mathfrak h}^{*2}$ 
is called {\it admissible} if $b=a+\bar{\varepsilon}_\mu$, 
for a certain $\mu\,(0\leqslant \mu \leqslant n-1)$. 
For $(a, b, c, d)\in {\mathfrak h}^{*4}$ let 
$\displaystyle W 
\left[ \left. \begin{array}{cc} 
c & d \\ b & a \end{array} 
\right| v \right] $ 
be the Boltzmann weight of the 
$A^{(1)}_{n-1}$ model for the state configuration 
$\displaystyle 
\left[ \begin{array}{cc} 
c & d \\ b & a \end{array} \right] $ 
round a face. 
Here the four states $a, b, c$ and $d$ are 
ordered clockwise from the SE corner. 
In this model $W 
\left[ \left. \begin{array}{cc} 
c & d \\ b & a \end{array} \right| 
v \right] =0~~$ 
unless the four pairs $(a,b), (a,d), (b,c)$ 
and $(d,c)$ are admissible. 
Non-zero Boltzmann weights are given by (\ref{eq:BW1}--\ref{eq:BW3}). 
See Appendix A. 

Among those, the weight (\ref{eq:BW2}) is different from 
the corresponding one used in our previous paper \cite{Bel-corr} 
by a minus sign. Accordingly, in the present paper 
we will use different definitions of 
the intertwining vectors (\ref{eq:int-vec}) 
and the type I vertex operators (\ref{eq:type-I}--\ref{eq:type-I'}) 
from the corresponding objects 
of \cite{Bel-corr} by extra factors of the form $(-1)^A$'s. 
This difference simply results from a gauge transformation. 

The Boltzmann weights 
solve the Yang-Baxter equation 
for the face model \cite{JMO}: 
\begin{equation}
\begin{array}{cc}
~ & 
\displaystyle \sum_{g} 
W\left[ \left. 
\begin{array}{cc} d & e \\ c & g \end{array} \right| 
v_1 \right]
W\left[ \left. 
\begin{array}{cc} c & g \\ b & a \end{array} \right| 
v_2 \right]
W\left[ \left. 
\begin{array}{cc} e & f \\ g & a \end{array} \right| 
v_1 -v_2 \right] \\
~ & ~ \\
= & \displaystyle \sum_{g} 
W\left[ \left. 
\begin{array}{cc} g & f \\ b & a \end{array} \right| 
v_1 \right]
W\left[ \left. 
\begin{array}{cc} d & e \\ g & f \end{array} \right| 
v_2 \right]
W\left[ \left. 
\begin{array}{cc} d & g \\ c & b \end{array} \right| 
v_1 -v_2 \right]
\end{array} \label{STR}
\end{equation}

\subsection{Vertex-face correspondence}

Let 
\begin{equation}
\begin{array}{rcl}
t(v)^a_{a-\bar{\varepsilon}_\mu}&=&
t(v; \epsilon , r)^a_{a-\bar{\varepsilon}_\mu}=
\displaystyle\sum_{\nu =0}^{n-1} \varepsilon_\nu 
t^\nu (v)^a_{a-\bar{\varepsilon}_\mu}, \\
t^\nu (v)^a_{a-\bar{\varepsilon}_\mu}&=&
\displaystyle\prod_{j=\mu +1}^{n-1} (-1)^{a_{\mu j}} 
\vartheta \left[\begin{array}{c} \frac{n}{2} \\ 
\frac{1}{2}+\frac{\nu}{n} \end{array} \right] 
\left( \frac{v}{nr}+\frac{\bar{a}_\mu}{r}; 
\frac{\pi \sqrt{-1}}{n\epsilon r} \right). \end{array}
\label{eq:int-vec}
\end{equation}
be the intertwining vectors. (See Appendix A, 
concerning the definition of $\bar{a}_\mu$.) 
Then $t(v)^a_{a-\bar{\varepsilon}_\mu}$'s relate 
the $R$-matrix of 
$(\mathbb{Z}/n\mathbb{Z})$-symmetric model in 
the principal regime and Boltzmann weights $W$ of 
$A^{(1)}_{n-1}$-model in the so-called regime III. (cf. figure 1) 
\begin{equation}
R(v_1-v_2)t (v_1)_a^d\otimes t (v_2)_d^c=
\sum_{b} t(v_1)_b^c \otimes t (v_2)_a^b 
W\left[ \left. 
\begin{array}{cc} c & d \\ b & a \end{array} \right| 
v_1 -v_2 \right]. 
\label{eq:Rtt=Wtt}
\end{equation}

\unitlength 1mm
\begin{picture}(100,20)
\put(23,0){
\begin{picture}(101,0)
\put(20,3){\begin{picture}(101,0)
\put(10,10){\vector(-1,0){10}}
\put(10,0){\vector(0,1){10}}
\put(8.8,5){\vector(-1,0){10}}
\put(8.,4.4){\scriptsize{$<$}}
\put(-4.5,4.2){$v_1$}
\put(4.15,8.5){\scriptsize{$\vee$}}
\put(5,8.8){\vector(0,-1){10}}
\put(3.9,-3.8){$v_2$}
\put(-2.5,10.5){$c$}
\put(10.5,-1.5){$a$}
\put(10.5,10.1){$d$}
\put(17,4){$=\;\displaystyle\sum_{b}$} 
\end{picture}
}
\put(58,3){\begin{picture}(101,0)
\put(10,0){\vector(-1,0){10}}
\put(10,0){\vector(0,1){10}}
\put(0,0){\vector(0,1){10}}
\put(10,10){\vector(-1,0){10}}
\put(-2.,4.4){\scriptsize{$<$}}
\multiput(0,5)(2.2,0){6}{\line(1,0){1.2}}
\put(-1.2,5){\vector(-1,0){2.5}}
\put(-7,4.2){$v_1$}
\put(4.15,-1.5){\scriptsize{$\vee$}}
\multiput(5,0)(0,2.2){6}{\line(0,1){1.2}}
\put(5,-1.2){\vector(0,-1){2.5}}
\put(3.9,-5.8){$v_2$}
\put(10.5,10.1){$d$}
\put(-2.5,10.5){$c$}
\put(10.5,-1.5){$a$}
\put(-2.5,-1.8){$b$}
\end{picture}
}
\end{picture}
}
\end{picture}

\vspace{2mm}

\begin{center}
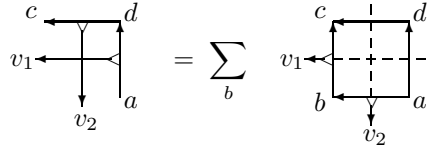
Figure 1. Picture representation of vertex-face 
correspondence. 
\end{center}

Note that the present intertwining vectors are different 
from the ones used in \cite{JMO}, which relate 
the $R$-matrix of 
$(\mathbb{Z}/n\mathbb{Z})$-symmetric model in 
the disordered phase and Boltzmann weights $W$ of 
$A^{(1)}_{n-1}$-model in the regime III. 

Let us introduce the dual intertwining vectors (see figure 2) 
satisfying 
\begin{equation}
\sum_{\mu =0}^{n-1} t_\mu^*  (v)^{a'}_{a}
t^\mu (v)^{a}_{a''} =\delta_{a''}^{a'}, ~~~~ 
\sum_{\nu =0}^{n-1} t^\mu (v)^{a}_{a-\bar{\varepsilon}_\nu} 
t_{\mu'}^* (v)^{a-\bar{\varepsilon}_\nu}_{a} =
\delta^\mu_{\mu'}. \label{eq:dual-t}
\end{equation}

\unitlength 1mm
\begin{picture}(100,20)
\put(40,3){\begin{picture}(101,0)
\put(-10,4){$\displaystyle\sum_{\mu=0}^{n-1}$}
\put(10.2,-3.){$a'$}
\put(-2.7,-2.8){$a$}
\put(4.15,0.1){\scriptsize{$\wedge$}}
\put(10,0){\vector(-1,0){10}}
\multiput(0,0)(0,2.2){5}{\line(0,1){1.2}}
\put(10.2,10.5){$a''$}
\put(-2.7,10.5){$a$}
\put(10,10){\vector(-1,0){10}}
\put(6,5){$\mu$}
\put(4.15,8.6){\scriptsize{$\vee$}}
\put(5,1.2){\line(0,1){7.6}}
\put(5,0){\line(0,-1){1}}
\put(5,-1.5){\vector(0,-1){2}}
\put(4.22,-6){$v$}
\put(15.,4.){$=\delta^{a''}_{a'}$,}
\end{picture}
}
\put(88,3){\begin{picture}(101,0)
\put(-10,4){$\displaystyle\sum_{a'}$}
\put(-3,4){$a$}
\put(11,4){$a'$}
\put(10,5){\vector(-1,0){10}}
\put(4.15,5.1){\scriptsize{$\wedge$}}
\put(4.15,3.6){\scriptsize{$\vee$}}
\put(5,6.4){\line(0,1){3.6}}
\put(6,7){$\mu'$}
\put(5,3.6){\vector(0,-1){3.6}}
\put(6,2){$\mu$}
\put(4.2,-2.2){$v$}
\put(17,4.){$=\delta^{\mu'}_{\mu}$.}
\end{picture}
}
\end{picture}

\vspace{3mm}

\begin{center}
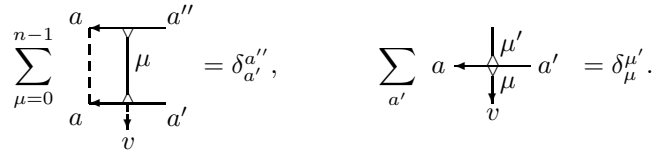
Figure 2. Picture representation of the dual intertwining 
vectors. 
\end{center}

{}From (\ref{eq:Rtt=Wtt}) and (\ref{eq:dual-t}), we have (cf. figure 3) 
\begin{equation}
t^*(v_{1})^{b}_{c}\otimes t^*(v_{2})^{a}_{b}
R(v_{1}-v_2 )=
\displaystyle\sum_{d} 
W\left[ \left. \begin{array}{cc} 
c & d \\ b & a \end{array} \right| v_{1}-v_2 \right]
t^*(v_{1} )^{a}_{d}\otimes t^*(v_{2} )^{d}_{c}. 
\label{eq:dJMO}
\end{equation}

\newpage

\unitlength 1mm
\begin{picture}(100,20)
\put(23,0){
\begin{picture}(101,0)
\put(20,3){\begin{picture}(101,0)
\put(10,0){\vector(-1,0){10}}
\put(0,0){\vector(0,1){10}}
\put(5,6.4){\line(0,1){3.6}}
\put(-.2,4.4){\scriptsize{$>$}}
\put(1.4,5){\line(1,0){10}}
\put(0,5){\line(-1,0){1}}
\put(-1.5,5){\vector(-1,0){2}}
\put(-6.5,4.2){$v_1$}
\put(4.15,.1){\scriptsize{$\wedge$}}
\put(5,1.4){\line(0,1){10}}
\put(5,0){\line(0,-1){1}}
\put(5,-1.5){\vector(0,-1){2}}
\put(4,-5.8){$v_2$}
\put(-2.5,10.5){$c$}
\put(10.5,-1.5){$a$}
\put(-2.5,-1.8){$b$}
\put(17,4){$=\;\displaystyle\sum_{d}$} 
\end{picture}
}
\put(56,3){\begin{picture}(101,0)
\put(10,0){\vector(-1,0){10}}
\put(10,0){\vector(0,1){10}}
\put(0,0){\vector(0,1){10}}
\put(10,10){\vector(-1,0){10}}
\put(9.8,4.4){\scriptsize{$>$}}
\multiput(10,5)(-2.2,0){6}{\line(-1,0){1.2}}
\put(11.6,5){\line(1,0){2}}
\put(-1.2,5){\vector(-1,0){2}}
\put(-7,4.2){$v_1$}
\put(4.15,10.1){\scriptsize{$\wedge$}}
\put(5,11.5){\line(0,1){2}}
\multiput(5,10)(0,-2.2){6}{\line(0,-1){1.2}}
\put(5,-1.2){\vector(0,-1){2}}
\put(4,-5.3){$v_2$}
\put(10.5,10.1){$d$}
\put(-2.5,10.5){$c$}
\put(10.5,-1.5){$a$}
\put(-2.5,-1.8){$b$}
\end{picture}
}
\end{picture}
}
\end{picture}

\vspace{3mm}

\begin{center}
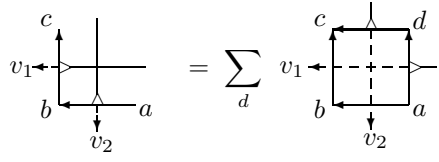
Figure 3. Vertex-face correspondence by dual intertwining 
vectors. 
\end{center}

For fixed $r>n-1$, let 
\begin{equation}
S(v )=-R(v)|_{r\mapsto r-1}, ~~~~
W'\left[ \left. 
\begin{array}{cc} c & d \\ b & a \end{array} \right| 
v \right]=-W\left[ \left. 
\begin{array}{cc} c & d \\ b & a \end{array} \right| 
v \right] \left. \makebox{\rule[-4mm]{0pt}{11mm}} 
\right|_{r\mapsto r-1}, 
\label{eq:SXYZ}
\end{equation}
and 
\begin{equation}
t'{}^* (u)^{b}_{a}:=t^* (u; \epsilon , r-1)^{b}_{a}. 
\label{eq:t'*}
\end{equation}
Then we have 
\begin{equation}
\displaystyle t'{}^*(v_1 )^{b}_{c}\otimes 
t'{}^*(v_2 )^{a}_{b}S(v_1 -v_2 ) 
=\displaystyle\sum_{d} 
W'\left[ \left. \begin{array}{cc} 
c & d \\ b & a \end{array} \right| v_1 -v_2 \right]
t'{}^*(v_1 )_{d}^{a}\otimes t'{}^*(v_2 )_{c}^{d}. 
\label{eq:sJMO}
\end{equation}

\section{Vertex operator algebra}

\subsection{Vertex operators for $(\mathbb{Z}/n\mathbb{Z})$-symmetric model}

Let ${\cal H}^{(i)}$ be the $\mathbb{C}$-vector space 
spanned by the half-infinite pure tensor vectors of the forms: 
\begin{equation}
\varepsilon_{\mu_1}\otimes \varepsilon_{\mu_2}\otimes 
\varepsilon_{\mu_3}\otimes \cdots 
~~~~ \mbox{with $\mu_j\in \mathbb{Z}/n\mathbb{Z}$, 
$\mu_j=i+1-j$ (mod $n$) for $j\gg 0$}. 
\label{eq:H^i}
\end{equation}
Let ${\cal H}^{*(i)}$ be the dual of ${\cal H}^{(i)}$ 
spanned by the half-infinite pure tensor vectors of the forms 
\begin{equation}
\cdots \otimes \varepsilon_{\mu_{-2}}\otimes \varepsilon_{\mu_{-1}}\otimes 
\varepsilon_{\mu_0}
~~~~ \mbox{with $\mu_j\in \mathbb{Z}/n\mathbb{Z}$, 
$\mu_j=i+1-j$ (mod $n$) for $j\ll 0$}. 
\end{equation}

Introduce the type I vertex operator by the following 
half-infinite transfer matrix 
\begin{equation}
\Phi^\mu (v_1 -v_2)=
\unitlength 0.5mm
\begin{picture}(100,20)
\put(15,-18){\begin{picture}(100,0)
\put(60,20){\vector(-1,0){60}}
\put(10,30){\vector(0,-1){20}}
\put(20,30){\vector(0,-1){20}}
\put(30,30){\vector(0,-1){20}}
\put(40,30){\vector(0,-1){20}}
\put(4,22){$\mu$}
\put(-7,18){$v_1$}
\put(9,5){$v_2$}
\put(19,5){$v_2$}
\put(29,5){$v_2$}
\put(39,5){$v_2$}
\put(45,22){$\cdots$}
\end{picture}
}
\end{picture}
\label{eq:B-I}
\end{equation}

~

\noindent Then the operator (\ref{eq:B-I}) is an intertwiner 
from ${\cal H}^{(i)}$ to ${\cal H}^{(i+1)}$. 
The type I vertex operators satisfy the following 
commutation relation: 
\begin{equation}
\Phi^\mu (v_1)\Phi^\nu (v_2)=
\sum_{\mu',\nu'} R(v_1-v_2)^{\mu\nu}_{\mu'\nu'} 
\Phi^{\nu'} (v_2)\Phi^{\mu'} (v_1). 
\label{eq:RPhiPhi}
\end{equation}

When we consider an operator related to `creation-annihilation' process, 
we need another type of vertex operators, the type II vertex operators 
that satisfy the following commutation relations: 
\begin{equation}
\Psi^*_\nu (v_2)\Psi^*_\mu (v_1)=
\sum_{\mu',\nu'} \Psi^*_{\mu'} (v_1)\Psi^*_{\nu'} (v_2)
S(v_1-v_2)_{\mu\nu}^{\mu'\nu'}, 
\label{eq:R'PsiPsi}
\end{equation}
\begin{equation}
\Phi^\mu (v_1)\Psi^*_\nu (v_2)=\chi (v_1 -v_2) 
\Psi^*_{\nu} (v_2)\Phi^{\mu} (v_1). 
\label{eq:chiPsiPhi}
\end{equation}

Let 
\begin{equation}
\rho^{(i)}=x^{2nH_{CTM}}: {\cal H}^{(i)}\rightarrow 
{\cal H}^{(i)}, 
\end{equation}
where $H_{CTM}$ is the CTM Hamiltoian defined in \cite{Bel-corr}. 
Then we have the homogeneity relation
\begin{equation}
\Phi^\mu (v) \rho^{(i)} =\rho^{(i+1)}\Phi^\mu (v-n), ~~~~ 
\Psi^*_\mu (v) \rho^{(i)} =\rho^{(i+1)}\Psi^*_\mu (v-n). 
\label{eq:homo}
\end{equation}

\subsection{Vertex operators for the $A^{(1)}_{n-1}$-model}

For $k=a+\rho , l=\xi +\rho$ and $0\leqslant i\leqslant n-1$, 
let ${\cal H}^{(i)}_{l,k}$ be the space of admissible paths 
$(a_0 , a_1, a_2, \cdots )$ such that 
\begin{equation}
a_0 =a, ~~~ a_{j} -a_{j+1}\in \left\{ 
\bar{\varepsilon}_0 , \bar{\varepsilon}_1 , 
\cdots , \bar{\varepsilon}_{n-1} 
\right\}, \mbox{ for $j=0, 1, 2, 3, \cdots$, }~~~~ 
a_j=\xi +\omega_{i+1-j} \mbox{ for 
$j\gg 0$}. 
\end{equation}
Also, let ${\cal H}^{*(i)}_{l,k}$ be the space of admissible paths 
$(\cdots , a_{-2} , a_{-1}, a_{0})$ such that 
\begin{equation}
a_0 =a, ~~~ a_{j} -a_{j+1}\in \left\{ 
\bar{\varepsilon}_0 , \bar{\varepsilon}_1 , 
\cdots , \bar{\varepsilon}_{n-1} 
\right\}, \mbox{ for $j=-1, -2, -3, \cdots$, }~~~~ 
a_j=\xi +\omega_{i+1-j} \mbox{ for 
$j\ll 0$}. 
\end{equation}
Introduce the type I vertex operator by the following 
half-infinite transfer matrix 
\begin{equation}
\hspace{2cm} \Phi(v_1 -v_2)_a^{a+\bar{\varepsilon}_\mu}=
\unitlength 1mm
\begin{picture}(100,10)
\put(-10,-60){\begin{picture}(101,0)
\put(20,55){\vector(0,1){10}}
\put(30,55){\vector(0,1){10}}
\put(30,55){\vector(-1,0){10}}
\put(30,65){\vector(-1,0){10}}
\put(40,55){\vector(0,1){10}}
\put(40,55){\vector(-1,0){10}}
\put(40,65){\vector(-1,0){10}}
\put(50,55){\vector(0,1){10}}
\put(50,55){\vector(-1,0){10}}
\put(50,65){\vector(-1,0){10}}
\put(60,55){\vector(-1,0){10}}
\put(60,65){\vector(-1,0){10}}
\put(19,53){$a$}
\put(16,67){$a\!+\!\bar{\varepsilon}_\mu$}
\multiput(18,60)(2,0){23}{\line(1,0){1}}
\put(17,60){\vector(-1,0){1.5}}
\multiput(25,53)(0,2){8}{\line(0,1){1}}
\multiput(35,53)(0,2){8}{\line(0,1){1}}
\multiput(45,53)(0,2){8}{\line(0,1){1}}
\put(25,52){\vector(0,-1){1.5}}
\put(35,52){\vector(0,-1){1.5}}
\put(45,52){\vector(0,-1){1.5}}
\put(24,48){$v_2$}
\put(34,48){$v_2$}
\put(44,48){$v_2$}
\put(12,59.5){$v_1$}
\end{picture}
}
\end{picture}
\label{eq:F-I}
\end{equation}

\vspace{7mm}

\noindent Then the operator (\ref{eq:F-I}) is an intertwiner 
from ${\cal H}^{(i)}_{l,k}$ to 
${\cal H}^{(i+1)}_{l,k+\bar{\varepsilon}_\mu}$. 
The type I vertex operators satisfy the following 
commutation relation: 
\begin{equation}
\Phi (v_1)^c_b\Phi (v_2)^b_a=
\sum_{d} W\left[ \left. \begin{array}{cc} 
c & d \\ 
b & a \end{array} \right| v_1-v_2 \right]
\Phi (v_2)^{c}_d\Phi (v_1)^{d}_a . 
\label{eq:Wphiphi}
\end{equation}
The free field realization of $\Phi (v_2)^b_a$ was constructed 
in \cite{AJMP}. See Sec 4.2. 

The type II vertex operators should satisfy 
the following commutation relations: 
\begin{equation}
\Psi^* (v_2)^{\xi_c}_{\xi_d}\Psi^* (v_1)^{\xi_d}_{\xi_a}=
\sum_{\xi_b} 
\Psi^* (v_1)^{\xi_c}_{\xi_b}\Psi^* (v_2)^{\xi_b}_{\xi_a} 
W'\left[ \left. \begin{array}{cc} 
\xi_c & \xi_d \\ 
\xi_b & \xi_a \end{array} \right| v_1-v_2 \right], 
\label{eq:W'psipsi}
\end{equation}
\begin{equation}
\Phi (v_1)^{a'}_a\Psi^* (v_2)^{\xi'}_{\xi}=
\chi (v_1-v_2)\Psi^* (v_2)^{\xi'}_{\xi}\Phi (v_1)^{a'}_a . 
\label{eq:Wchipsiphi}
\end{equation}

Let 
\begin{equation}
\rho^{(i)}_{l,k}=G_a x^{2nH_{l,k}^{(i)}}, 
\label{eq:rho_lk}
\end{equation}
where 
$$
G_a =\prod_{0\leqslant\mu <\nu\leqslant n-1} [a_{\mu\nu}]. 
$$
Then we have the homogeneity relation
\begin{equation}
\Phi (v)^{a'}_a \dfrac{\rho^{(i)}_{a+\rho , l}}{G_a} 
=\dfrac{\rho^{(i+1)}_{a'+\rho , l}}{G_{a'}}\Phi (v-n)^{a'}_a, 
~~~~ 
\Psi^* (v)^{\xi'}_{\xi} \rho^{(i)}_{k,\xi +\rho} =
\rho^{(i+1)}_{k,\xi' +\rho}\Psi^* (v-n)^{\xi'}_{\xi}. 
\label{eq:Whomo}
\end{equation}
The free field realization of $\Psi^* (v)^{\xi'}_{\xi}$ was constructed 
in \cite{FHSY,FKQ}. See Sec 4.3.

\subsection{Tail operators and commutation relations}

In \cite{Bel-corr} we introduced the intertwining operators between 
${\cal H}^{(i)}$ and 
${\cal H}^{(i)}_{l,k}$ ($k=l+\omega_{i}$ (mod $Q$)): 
\begin{equation}
\begin{array}{rcl}
T(u ){}^{\xi a_0}&=&
\displaystyle\prod_{j=0}^\infty 
t^{\mu_j}(-u ){}^{a_j}_{a_{j+1}}: 
{\cal H}^{(i)}\rightarrow {\cal H}^{(i)}_{l,k}, \\
T(u ){}_{\xi a_0}&=&
\displaystyle\prod_{j=0}^\infty 
t^*_{\mu_j}(-u ){}_{a_j}^{a_{j+1}}: 
{\cal H}^{(i)}_{l,k}\rightarrow {\cal H}^{(i)}, 
\end{array}
\label{eq:T^_}
\end{equation}
which satisfy 
\begin{equation}
\rho^{(i)}=\left( \dfrac{(x^{2r-2};x^{2r-2})_\infty}{
(x^{2r};x^{2r})_\infty} \right)^{(n-1)(n-2)/2}\dfrac{1}{
G'_\xi} \sum_{k\equiv l+\omega_i\atop\mbox{\scriptsize (mod $Q$)}} 
T (u)_{a\xi} \rho^{(i)}_{l,k} T(u)^{a\xi}, 
\label{eq:rho-rel}
\end{equation}
and the intertwining relations 
\begin{equation}
T(u)^{\xi b} \Phi^\mu (v) =\sum_a t^\mu 
(v-u ){}_{a}^{b}\Phi (v)^b_aT(u)^{\xi a}, 
\label{eq:T^Phi}
\end{equation}
\begin{equation}
T(u)_{\xi b} \Phi (v)^b_a =\sum_\mu t^*_\mu 
(v-u ){}^{a}_{b}\Phi^\mu (v)T(u)_{\xi a}. 
\label{eq:T_phi}
\end{equation}
Here, $k=a_0 +\rho$ and $l=\xi +\rho$, and 
$0<\Re (u)<\frac{n}{2}+1$. 

In order to obtain the form factors of 
$(\mathbb{Z}/n\mathbb{Z})$-symmetric model, we need 
the free field representations of the tail operator 
which is offdiagonal with respect to the boundary conditions 
(see figure 4): 
\begin{equation}
\Lambda (u )_{\xi\,a}^{\xi'a'}=T(u )^{\xi' a'}T(u)_{\xi\, a}: 
{\cal H}^{(i)}_{l,k}\rightarrow {\cal H}^{(i)}_{l'k'}, 
\label{eq:L=TT'}
\end{equation}
where $k=a+\rho$, $l=\xi +\rho$, $k'=a'+\rho$, and $l'=\xi' +\rho$. 
Let 
\begin{equation}
L\left[  \left. \begin{array}{cc} a'_0 & a'_1 \\
a_0 & a_1 \end{array} \right| u \right] :=
\sum_{\mu =0}^{n-1} t^*_\mu (-u)_{a_0}^{a_1} 
t^\mu (-u)^{a'_0}_{a'_1}. 
\label{eq:Lop}
\end{equation}
Then we have 
\begin{equation}
\Lambda(u ){}_{\xi\,a_0}^{\xi'a'_0}=
\prod_{j=0}^\infty L\left[  \left. \begin{array}{cc} 
a'_j & a'_{j+1} \\
a_j & a_{j+1} \end{array} \right| u \right]. 
\label{eq:Lambda}
\end{equation}
\vspace{5mm}

\unitlength 1.4mm
\begin{picture}(100,20)
\put(-18,0){\begin{picture}(101,0)
\put(18,10){$\Lambda(u ){}_{\xi\,a_0}^{\xi'a'_0}=$}
\multiput(65,5)(-10,0){3}{\vector(-1,0){10}}
\multiput(66,5)(2,0){5}{\line(1,0){1}}
\multiput(65,15)(-10,0){3}{\vector(-1,0){10}}
\multiput(66,15)(2,0){5}{\line(1,0){1}}
\put(110,5){\vector(-1,0){10}}
\put(110,15){\vector(-1,0){10}}
\put(100,5){\vector(-1,0){10}}
\put(100,15){\vector(-1,0){10}}
\put(90,5){\line(-1,0){4}}
\put(90,15){\line(-1,0){4}}
\put(86,5){\vector(-1,0){10}}
\put(86,15){\vector(-1,0){10}}
\put(34,2.3){$a_0$}
\put(44,2.3){$a_1$}
\put(54,2.3){$a_2$}
\put(64,2.3){$a_3$}
\put(34,16.){$a'_0$}
\put(44,16.2){$a'_1$}
\put(54,16.2){$a'_2$}
\put(64,16.2){$a'_3$}
\put(75,2.3){$\xi$}
\put(84,2.3){$\cdots$}
\put(89,2.3){$\xi\!\!+\!\!\omega_2$}
\put(99,2.3){$\xi\!\!+\!\!\omega_1$}
\put(109,2.3){$\xi$}
\put(75,16.){$\xi'$}
\put(84,16){$\cdots$}
\put(89,16.){$\xi'\!\!+\!\!\omega_2$}
\put(99,16){$\xi'\!\!+\!\!\omega_1$}
\put(109,16.){$\xi'$}
\put(39.4165,5.){\scriptsize{$\wedge$}}
\put(39.4165,13.9){\scriptsize{$\vee$}}
\put(40,6.0){\line(0,1){7.9}}
\put(40,5){\line(0,-1){1}}
\put(40,3.5){\vector(0,-1){2}}
\put(38.,-.3){$-u$}
\put(49.4165,5.){\scriptsize{$\wedge$}}
\put(49.4165,13.9){\scriptsize{$\vee$}}
\put(50,6.){\line(0,1){7.9}}
\put(59.4165,5.){\scriptsize{$\wedge$}}
\put(59.4165,13.9){\scriptsize{$\vee$}}
\put(60,6.){\line(0,1){7.9}}
\put(80.4165,5.){\scriptsize{$\wedge$}}
\put(80.4165,13.9){\scriptsize{$\vee$}}
\put(81,6.){\line(0,1){7.9}}
\put(94.4165,5.){\scriptsize{$\wedge$}}
\put(94.4165,13.9){\scriptsize{$\vee$}}
\put(95,6.){\line(0,1){7.9}}
\put(104.4165,5.){\scriptsize{$\wedge$}}
\put(104.4165,13.9){\scriptsize{$\vee$}}
\put(105,6.){\line(0,1){7.9}}
\end{picture}
}
\end{picture}

\vspace{3mm}

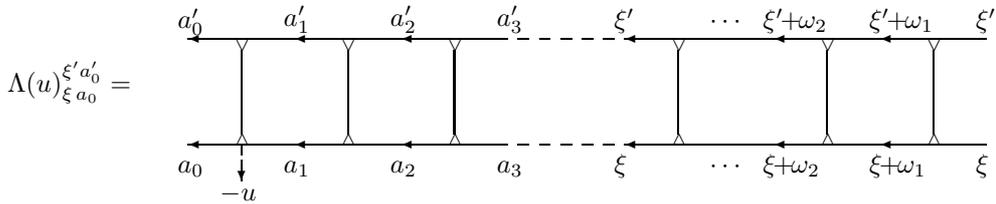
Figure 4. Tail operator $\Lambda(u ){}^{\xi'a'_0}_{\xi\,a_0}$. 
The upper (resp. lower) half stands for $T(u ){}^{\xi' a'_0}$ (resp. 
$T(u ){}_{\xi\,a_0}$). 

\vspace{7mm}

Note that 
\begin{equation}
L\left[ \left. \begin{array}{cc} a' & a'-\bar{\varepsilon}_\nu \\
a & a-\bar{\varepsilon}_\mu \end{array} \right| u \right] = 
\dfrac{[u+\bar{a}_\mu -\bar{a'}_\nu ]}{[u]}
\prod_{j\neq\mu} \dfrac{[\bar{a'}_\nu -\bar{a}_{j}]}
{[a_{\mu j}]}. 
\label{eq:Lop-ex}
\end{equation}
It is obvious from (\ref{eq:dual-t}), we have 
\begin{equation}
L\left[ \left. \begin{array}{cc} a & a' \\
a & a'' \end{array} \right| u \right] = \delta_{a''}^{a'}.  
\label{eq:L-inv}
\end{equation}
We therefore have 
\begin{equation}
\Lambda (u)^{\xi' a}_{\xi\,a}=\delta^{\xi'}_\xi . \label{eq:Lambda=1}
\end{equation}

From (\ref{eq:T^Phi}), (\ref{eq:T_phi}) and the definition of 
the tail operator (\ref{eq:L=TT'}) we have 
\begin{equation}
\Lambda (u )^{\xi'c}_{\xi\,b}\Phi (v)^b_a=
\sum_{d}L\left[ \left. \begin{array}{cc} c & d \\
b & a \end{array} \right| u-v \right]
\Phi (v)^c_d\Lambda (u)^{\xi'd}_{\xi\,a}. 
\label{eq:Lambda-phi}
\end{equation}

Consider the algebra 
\begin{equation}
\Psi^* (v)^{\xi'}_\xi T(u)^{\xi a}  =\sum_\mu 
T(u)^{\xi' a}\Psi^*_\mu (v)
t'{}^\mu (v-u-\Delta u ){}^{\xi'}_{\xi}, 
\label{eq:T^psi}
\end{equation}
\begin{equation}
\Psi^{*}_{\mu} (v)T(u)_{\xi a}  =\sum_{\xi'} 
T(u)_{\xi' a}\Psi^* (v)^{\xi'}_\xi 
t'{}^*_\mu (v-u-\Delta u ){}_{\xi'}^{\xi}. 
\label{eq:T_Psi}
\end{equation}
From these, we have 
\begin{equation}
\Psi^* (v)^{\xi_c}_{\xi_d}\Lambda (u )^{\xi_d\,a'}_{\xi_a\,a}=
\sum_{\xi_b}L'\left[ \left. \begin{array}{cc} \xi_c & \xi_d \\
\xi_b & \xi_a \end{array} \right| u+\Delta u-v \right]
\Lambda (u)^{\xi_c\,a'}_{\xi_b\,a}\Psi^* (v)^{\xi_b}_{\xi_a}, 
\label{eq:Lambda-psi}
\end{equation}
where 
\begin{equation}
L'\left[ \left. \begin{array}{cc} \xi_c & \xi_d \\
\xi_b & \xi_a \end{array} \right| u\right]
=L\left. \left[ \left. \begin{array}{cc} \xi_c & \xi_d \\
\xi_b & \xi_a \end{array} \right| u\right]\right|_{r\mapsto r-1}. 
\end{equation}
We should find a representation of $\Lambda (u )^{\xi'a'}_{\xi\,a}$ 
and fix the constant $\Delta u$ 
that solves (\ref{eq:Lambda-phi}) and (\ref{eq:Lambda-psi}). 

\section{Free filed realization}

One of the most standard ways to calculate correlation functions 
and form foctors 
is the vertex operator approach \cite{JMbk} 
on the basis of free field representation. 
The free field representations for the type I vertex operators 
of the $A^{(1)}_{n-1}$ model were constructed in \cite{AJMP}, 
in terms of oscillators introduced in \cite{FL,AKOS}. 
Those for the type II vertex operators were constructed in 
\cite{FHSY,FKQ}, also 
in terms of oscillators introduced in \cite{FL,AKOS}. 
It was shown in \cite{KK1,KK2} that the elliptic algebra 
$U_{q,p}(\widehat{\mathfrak s\mathfrak l}_N)$ provides 
the Drinfeld realization of the face type elliptic quantum group 
${\cal B}_{q,\lambda}(\widehat{\mathfrak s\mathfrak l}_N)$ tensored by 
a Heisenberg algebra. 
Using these representations we derive the free field representation 
of the tail operator in this section. 

\subsection{Bosons}

Let us consider the bosons
$B_m^j\,(1\leqslant j \leqslant n-1, m \in \mathbb{Z}
\backslash \{0\})$
with the commutation relations
\begin{equation}
[B_m^j,B_{m'}^k]
=\left\{ \begin{array}{ll} 
m\dfrac{[(n-1)m]_x}{[nm]_x}
\dfrac{[(r-1)m]_x}{[rm]_x}\delta_{m+m',0}, & (j=k)\\
-mx^{{\rm sgn}(j-k)nm}\dfrac{[m]_x}{[nm]_x}
\dfrac{[(r-1)m]_x}{[rm]_x}\delta_{m+m',0}, & (j\neq k), 
\end{array} \right. 
\label{eq:comm-B}
\end{equation}
where the symbol $[a]_x$ stands for
$(x^a-x^{-a})/(x-x^{-1})$.
Define $B_m^n$ by
\begin{eqnarray*}
\sum_{j=1}^n x^{-2jm}B_m^j=0.
\end{eqnarray*}
Then the commutation relations (\ref{eq:comm-B}) 
holds for all $1\leqslant j,k \leqslant n$.
These oscillators were introduced in \cite{FL,AKOS}. 

For $\alpha, \beta \in {\mathfrak h}^*$ let us define 
the zero mode operators $P_\alpha, Q_\beta$ 
with the commutation relations 
\begin{equation*}
[P_{\alpha},\sqrt{-1}Q_{\beta}]=\langle \alpha,\beta \rangle, 
~~~~ [P_{\alpha}, B_m^j ]=
[Q_{\beta}, B_m^j ]=0. 
\end{equation*}

~\\
We will deal with the bosonic Fock spaces 
${\cal{F}}_{l,k}, (l,k \in {\mathfrak h}^*)$
generated by $B_{-m}^j (m>0)$
over the vacuum vectors $|l,k\rangle$ :
\begin{eqnarray*}
{\cal{F}}_{l,k}=
\mathbb{C}[\{ B_{-1}^j, B_{-2}^j,\cdots \}_{
1\leqslant j \leqslant n}]|l,k\rangle,
\end{eqnarray*}
where
\begin{eqnarray*}
B_m^j|l,k\rangle&=&0 ~(m>0),\\
P_{\alpha}|l,k\rangle &=&\langle \alpha,
\beta_1 k+\beta_2 l \rangle
|l,k\rangle,\\
|l,k\rangle&=&\exp \left(\sqrt{-1}(\beta_1Q_k+
\beta_2Q_l)\right)|0,0\rangle, 
\end{eqnarray*}
where $\beta_1$ and $\beta_2$ are defined by 
\begin{equation}
t^2 -\beta_0 t-1=(t-\beta_1)(t-\beta_2), ~~~~ 
\beta_0 =\dfrac{1}{\sqrt{r(r-1)}}, ~~~~ \beta_1<\beta_2, 
\label{eq:beta_12}
\end{equation}

\subsection{Type I vertex operators}

Let us define the basic operators for $j=1,\cdots,n-1$
\begin{eqnarray}
U_{-\alpha_j}(v)&=&
\exp\left(-\beta_1 (\sqrt{-1}Q_{\alpha_j}
+P_{\alpha_j}\log  z)\right) 
:\exp\left(\sum_{m \neq 0}\frac{1}{m}
(B_m^j-B_m^{j+1})(x^jz)^{-m}\right):, \\
U_{\omega_j}(v)&=&
\exp\left(\beta_1 (\sqrt{-1}Q_{\omega_j}
+P_{\omega_j}\log z)\right)
:\exp\left(-\sum_{m\neq 0}\frac{1}{m} \sum_{k=1}^j 
x^{(j-2k+1)m}B_m^kz^{-m}\right):, 
\end{eqnarray}
where $\beta_1 =-\sqrt{\frac{r-1}{r}}$ and $z=x^{2v}$ as usual. 
For some useful OPE formulae and commutation relations, 
see Appendix B. 

In the sequel we set
\begin{eqnarray*}
\pi_\mu=\sqrt{r(r-1)}P_{\bar{\varepsilon}_\mu},~
\pi_{\mu \nu}=\pi_\mu-\pi_\nu =rL_{\mu\nu}-(r-1)K_{\mu\nu}.
\end{eqnarray*}
The operators $K_{\mu\nu}$, $L_{\mu\nu}$ and $\pi_{\mu \nu}$ 
act on ${\cal{F}}_{l,k}$ as scalors 
$\langle \varepsilon_\mu -\varepsilon_\nu, k\rangle$, 
$\langle \varepsilon_\mu -\varepsilon_\nu, l\rangle$ and 
$\langle \varepsilon_\mu -\varepsilon_\nu, rl-(r-1)k\rangle$, 
respectively. In what follows we often use the symbols 
$$
G_K =\prod_{0\leqslant\mu <\nu\leqslant n-1} [K_{\mu\nu}], ~~~~
G'_L =\prod_{0\leqslant\mu <\nu\leqslant n-1} [L_{\mu\nu}]'. 
$$

For $0 \leqslant \mu \leqslant n-1$ 
define the type I vertex operator \cite{AJMP} by
\begin{eqnarray}
&&\phi_\mu (v_0 )=\displaystyle\oint_C 
\prod_{j=1}^{\mu}\frac{dz_j}{2\pi \sqrt{-1} z_j}
U_{\omega_1}(v_0 )U_{-\alpha_1}(v_1)\cdots 
U_{-\alpha_\mu}(v_{\mu})
\prod_{j=0}^{\mu-1}f(v_{j+1}-v_{j},K_{j \mu}) 
\prod_{j=0\atop j\neq\mu}^{n-1} [K_{j\mu}]^{-1} 
\label{eq:type-I} \\
&=&(-1)^\mu \displaystyle\oint_C 
\prod_{j=1}^{\mu}\frac{dz_j}{2\pi \sqrt{-1} z_j}
U_{-\alpha_\mu}(v_{\mu})\cdots U_{-\alpha_1}(v_1)
U_{\omega_1}(v_0 )
\prod_{j=0}^{\mu-1}f(v_{j}-v_{j+1}, 1-K_{j \mu}) 
\prod_{j=0\atop j\neq\mu}^{n-1} [K_{j\mu}]^{-1}, 
\label{eq:type-I'} 
\end{eqnarray}
where $z_j=x^{2v_j}$. 
Considering the factors 
$f(v_{j+1}-v_{j}, K_{j \mu})$'s together with the OPE 
formulae (\ref{eq:OjAj-prod}) and (\ref{eq:AjAj+1-prod}), 
the expressions (\ref{eq:type-I}) has poles at 
$z_j=x^{\pm (1+2kr)}z_{j-1}\,(k 
\in \mathbb{Z}_{\geqslant 0})$. 
The integral contour $C$ for $z_j$-integration should be 
chosen such that all integral variables lie in the 
common convergence domain; i.e., the contour $C$ 
encircles the poles at $z_j=x^{1+2kr}z_{j-1}\,(k 
\in \mathbb{Z}_{\geqslant 0})$, but not the poles at 
$z_j=x^{-1-2kr}z_{j-1}\,(k \in 
\mathbb{Z}_{\geqslant 0})$. 

Note that 
\begin{equation}
\phi_\mu(v): {\cal{F}}_{l,k} 
\longrightarrow {\cal{F}}_{l,k+\bar{\varepsilon}_\mu}. 
\label{eq:k-shift}
\end{equation}
These type I vertex operators satisfy 
the following commutation relations on 
${\cal{F}}_{l,k}$: 
\begin{eqnarray}
\phi_{\mu_1}(v_1)\phi_{\mu_2}(v_2)
=\sum_{\varepsilon_{\mu_1}+\varepsilon_{\mu_2}
=\varepsilon_{\mu_1'}+\varepsilon_{\mu_2'} }W\left[\left.
\begin{array}{cc}
a+\bar{\varepsilon}_{\mu_1}+\bar{\varepsilon}_{\mu_2}&
a+\bar{\varepsilon}_{\mu_1'}\\
a+\bar{\varepsilon}_{\mu_2}&a
\end{array}\right|v_1-v_2 \right]
{\phi}_{\mu_2'}(v_2)
{\phi}_{\mu_1'}(v_1). 
\label{eq:CR-I}
\end{eqnarray}
We thus denote the operator $\phi_\mu (v)$ by 
$\Phi (v )^{a+\bar{\varepsilon}_\mu}_a$ 
on the bosonic Fock space ${\cal{F}}_{l,a+\rho}$. 

Dual vertex operators are likewise defined as follows: 
\begin{equation}
\begin{array}{rcl}
\phi^*_\mu(v)&=&(-1)^{n-1-\mu} c_n^{-1}\displaystyle\oint
\prod_{j=\mu +1}^{n-1}\frac{dz_j}{2\pi \sqrt{-1} z_j}
U_{\omega_{n-1}}\left(v-\frac{n}{2} \right)U_{-\alpha_{n-1}}(v_{n-1})\cdots 
U_{-\alpha_{\mu +1}}(v_{\mu +1}) \\
&\times&\displaystyle\prod_{j=\mu +1}^{n-1}f(v_{j}-v_{j+1},K_{\mu j}) \\
&=&c_n^{-1}\displaystyle \oint
\prod_{j=\mu +1}^{n-1}\frac{dz_j}{2\pi \sqrt{-1} z_j}
U_{-\alpha_{\mu +1}}(v_{\mu +1})\cdots U_{-\alpha_{n-1}}(v_{n-1})
U_{\omega_{n-1}}\left(v-\frac{n}{2}\right) \\
&\times&\displaystyle\prod_{j=\mu +1}^{n-1}f(v_{j+1}-v_{j},1-K_{\mu j}). 
\label{eq:type-I*}
\end{array}
\end{equation}
Here $v_n =v-\tfrac{n}{2}$, and 
$$
c_n =x^{\frac{r-1}{r}\frac{n-1}{2n}} \dfrac{g_{n-1}(x^n)}{
(x^2; x^{2r})_\infty^n (x^{2r};x^{2r})_\infty^{2n-3}}, 
$$ 
where $g_{n-1} (z)$ is defined by (\ref{eq:g-def}). 
The integral contour for $z_j$-integration 
encircles the poles at $z_j=x^{1+2kr}z_{j+1}\,(k 
\in \mathbb{Z}_{\geqslant 0})$, but not the poles at 
$z_j=x^{-1-2kr}z_{j+1}\,(k \in 
\mathbb{Z}_{\geqslant 0})$, for $\mu +1\leqslant j\leqslant n-1$. 
Note that 
\begin{equation}
\phi^*_\mu (v): {\cal{F}}_{l,k} 
\longrightarrow {\cal{F}}_{l,k-\bar{\varepsilon}_\mu}. 
\label{eq:k-shift*}
\end{equation}
The operators $\phi_\mu (v)$ and $\phi^*_\mu (v)$ are dual 
in the following sense \cite{AJMP}: 
\begin{equation}
\sum_{\mu =0}^{n-1} \phi^*_\mu (v)\phi_\mu (v)=1. 
\label{eq:dualrel}
\end{equation}

In \cite{Bel-corr} we obtained the free field representation of 
$\Lambda (u)_{\xi\,a}^{\xi\,a'}$ satisfying 
(\ref{eq:Lambda-phi}) for $\xi'=\xi$: 
\begin{equation}
\begin{array}{rcl}
\Lambda (u)^{\xi\,a-\bar{\varepsilon}_{\mu}}_{\xi\,a-\bar{\varepsilon}_{\nu}}
&=&\displaystyle G_K
\oint \prod_{j=\mu +1}^{\nu} 
\dfrac{dz_j}{2\pi\sqrt{-1}z_j} 
U_{-\alpha_{\mu +1}}(v_{\mu +1}) \cdots 
U_{-\alpha_{\nu}}(v_{\nu}) \\
&\times& \displaystyle\prod_{j=\mu}^{\nu -1} 
f(v_{j+1}-v_j, K_{j\nu}) 
G_K^{-1}, 
\end{array}
\label{eq:Bose-Lambda}
\end{equation}
where $v_\mu =u$ and $\mu <\nu$.

\subsection{Type II vertex operators}

Let us define the basic operators for $j=1,\cdots,n-1$
\begin{eqnarray}
V_{-\alpha_j}(v)&=&
\exp\left(-\beta_2 (\sqrt{-1}Q_{\alpha_j}
+P_{\alpha_j}\log  z)\right)
:\exp\left(-\sum_{m \neq 0}\frac{1}{m}
(A_m^j-A_m^{j+1})(x^jz)^{-m}\right):, \\
V_{\omega_j}(v)&=&
\exp\left(\beta_2 (\sqrt{-1}Q_{\omega_j}
+P_{\omega_j}\log z\right)
:\exp\left(\sum_{m\neq 0}\frac{1}{m} \sum_{k=1}^j 
x^{(j-2k+1)m}A_m^kz^{-m}\right):, 
\end{eqnarray}
where $\beta_2 =\sqrt{\frac{r}{r-1}}$ and $z=x^{2v}$, and 
\begin{equation}
A_m^j=(-1)^m\dfrac{[rm]_x}{[(r-1)m]_x}B_m^j . 
\end{equation}
For some useful OPE formulae and commutation relations, see 
Appendix B. 

For $0 \leqslant \mu \leqslant n-1$ 
define the type II vertex operator 
\cite{FHSY,FKQ}\footnote{
Precisely speaking, the integral contour for $\psi^*_\mu(v_0 )$ 
of \cite{FHSY} is different from that of \cite{FKQ}. The contour 
should be chosen in such a way that all integral variables lie in 
the convergence domain of the integral formula 
(\ref{eq:type-II}). In the present paper we adopt the contour of 
\cite{FKQ}. } by
\begin{eqnarray}
\psi^*_\mu(v_0 )&=&\displaystyle\oint_{C'} 
\prod_{j=1}^{\mu}\frac{dz_j}{2\pi \sqrt{-1} z_j}
V_{\omega_1}(v_0 )V_{-\alpha_1}(v_1)\cdots 
V_{-\alpha_\mu}(v_{\mu})
\prod_{j=0}^{\mu-1}f^*(v_{j+1}-v_{j},L_{j \mu}) 
\label{eq:type-II} \\
&=&(-1)^\mu \displaystyle\oint_{C'} 
\prod_{j=1}^{\mu}\frac{dz_j}{2\pi \sqrt{-1} z_j}
V_{-\alpha_\mu}(v_{\mu})\cdots V_{-\alpha_1}(v_1)
V_{\omega_1}(v_0 )
\prod_{j=0}^{\mu-1}f^*(v_{j}-v_{j+1}, 1-L_{j \mu}), 
\label{eq:type-II'} 
\end{eqnarray}
where $z_j=x^{2v_j}$. Considering the factors 
$f^*(v_{j+1}-v_{j}, L_{j \mu})$'s together with the OPE 
formulae (\ref{eq:OjAj-prod'}) and (\ref{eq:AjAj+1-prod'}), 
the expressions (\ref{eq:type-II}) has poles at 
$z_j=x^{\pm (-1+2k(r-1))}z_{j-1}\,(k 
\in \mathbb{Z}_{\geqslant 0})$. 
The integral contour $C'$ for $z_j$-integration should be chosen such that 
$C'$ encircles the poles at $z_j=x^{-1+2k(r-1)}z_{j-1}\,(k 
\in \mathbb{Z}_{\geqslant 0})$, but not the poles at 
$z_j=x^{1-2k(r-1)}z_{j-1}\,(k \in 
\mathbb{Z}_{\geqslant 0})$. 

Note that 
\begin{equation}
\psi^*_\mu(v): {\cal{F}}_{l,k} 
\longrightarrow {\cal{F}}_{l+\bar{\varepsilon}_\mu,k}. 
\label{eq:l-shift}
\end{equation}
These type II vertex operators satisfy 
the following commutation relations on 
${\cal{F}}_{l,k}$: 
\begin{eqnarray}
\psi^*_{\mu_1}(v_1)\psi^*_{\mu_2}(v_2)
=\sum_{\varepsilon_{\mu_1}+\varepsilon_{\mu_2}
=\varepsilon_{\mu_1'}+\varepsilon_{\mu_2'} }W'\left[\left.
\begin{array}{cc}
\xi +\bar{\varepsilon}_{\mu_1}+\bar{\varepsilon}_{\mu_2}&
\xi +\bar{\varepsilon}_{\mu_2}\\
\xi +\bar{\varepsilon}_{\mu_1'}&\xi 
\end{array}\right|v_2-v_1 \right]
{\psi^*}_{\mu_2'}(v_2)
{\psi^*}_{\mu_1'}(v_1). 
\label{eq:CR-II}
\end{eqnarray}
We thus denote the operator $\psi^*_\mu (v)$ by 
$\Psi^* (v )^{\xi+\bar{\varepsilon}_\mu}_\xi$ 
on the bosonic Fock space ${\cal{F}}_{\xi +\rho,k}$. 

Dual vertex operators are likewise defined as follows: 
\begin{equation}
\begin{array}{rcl}
\psi_\mu(v)&=&(-1)^{n-1-\mu}c'_n{}^{-1} \displaystyle\oint
\prod_{j=\mu +1}^{n-1}\frac{dz_j}{2\pi \sqrt{-1} z_j}
V_{\omega_{n-1}}\left(v-\frac{n}{2} \right)V_{-\alpha_{n-1}}(v_{n-1})\cdots 
V_{-\alpha_{\mu +1}}(v_{\mu +1}) \\
&\times&\displaystyle\prod_{j=\mu +1}^{n-1}
f^*(v_{j}-v_{j+1},L_{\mu j}) \prod_{j=0\atop j\neq\mu}^{n-1} 
\dfrac{[1]'}{[L_{j\mu}]'} \\
&=&c'_n{}^{-1} \displaystyle \oint
\prod_{j=\mu +1}^{n-1}\frac{dz_j}{2\pi \sqrt{-1} z_j}
V_{-\alpha_{\mu +1}}(v_{\mu +1})\cdots V_{-\alpha_{n-1}}(v_{n-1})
V_{\omega_{n-1}}\left(v-\frac{n}{2}\right) \\
&\times&\displaystyle\prod_{j=\mu +1}^{n-1}
f^*(v_{j+1}-v_{j},1-L_{\mu j})\prod_{j=0\atop j\neq\mu}^{n-1} 
\dfrac{[1]'}{[L_{j\mu}]'}. 
\label{eq:type-II*}
\end{array}
\end{equation}
Here $v_n =v-\tfrac{n}{2}$, and 
$$
c'_n=x^{\frac{r(n-1)}{2(r-1)}}
\left(\frac{(x^{2r};x^{2r-2})_\infty}
{(x^{2r-2};x^{2r-2})_\infty}\right)^n 
g_{n-1}^*(x^n), 
$$
where $g_{n-1}^* (z)$ is defined by (\ref{eq:g*-def}). 
The integral contour for $z_j$-integration encircles 
the poles at $z_j=x^{-1+2k(r-1)}z_{j+1}\,
(k \in \mathbb{Z}_{\geqq 0})$, but not 
the poles at $z_j=x^{1-2k(r-1)}z_{j+1}\,(k \in 
\mathbb{Z}_{\geqq 0})$, for $\mu +1\leqslant j\leqslant n-1$. 
Note that 
\begin{equation}
\psi_\mu (v): {\cal{F}}_{l,k} 
\longrightarrow {\cal{F}}_{l-\bar{\varepsilon}_\mu,k}. 
\label{eq:l-shift*}
\end{equation}
The operators $\psi_\mu (v)$ and $\psi^*_\mu (v)$ are dual 
in the following sense \cite{FKQ}: 
\begin{equation}
\psi_\mu (v)\psi^*_\nu (v')=\delta_{\mu\nu}
\frac{1}{1-z'/z}+
\mbox{(regular terms at $v=v'$)}. 
\label{eq:dualrel*}
\end{equation}

For later convenience, we also introduce another type of 
basic operators: 
\begin{equation}
W_{-\alpha_j} (v)=
\exp\left(-\beta_0 (\sqrt{-1}Q_{\alpha_j}
+P_{\alpha_j}\log  (-1)^rz)\right)
:\exp\left(-\sum_{m \neq 0}\frac{1}{m}
(O_m^j-O_m^{j+1})(x^jz)^{-m}\right):, 
\label{eq:IxII}
\end{equation}
where $\beta_0 =\beta_1 +\beta_2 =\dfrac{1}{\sqrt{r(r-1)}}$, 
$(-1)^r:=\exp (\pi\sqrt{-1}r)$ and 
\begin{equation}
O_m^j =\dfrac{[m]_x}{[(r-1)m]_x} B_m^j . 
\end{equation}
Concerning useful OPE formulae and commutation relations, 
see Appendix B. 

\subsection{Free field realization of tail operators}

Consider (\ref{eq:Lambda-psi}) for $(\xi_a, \xi_d, \xi_c)=
(\xi ,\xi , \xi+\bar{\varepsilon}_{n-1})$, and 
$(a,a')\rightarrow (a+\bar{\varepsilon}_{n-2}, 
a+\bar{\varepsilon}_{n-1})$: 
\begin{equation}
\Psi^* (v)_\xi^{\xi +\bar{\varepsilon}_{n-1}} 
\Lambda (u)^{\xi\,a+\bar{\varepsilon}_{n-1}}_{\xi\,a+\bar{\varepsilon}_{n-2}}
=\sum_{\mu =0}^{n-1} L'\left[ \left. \begin{array}{cc} 
\xi +\bar{\varepsilon}_{n-1} & \xi \\
\xi +\bar{\varepsilon}_{\mu} & \xi \end{array} \right| u+\Delta u-v \right]
\Lambda (v)^{\xi +\bar{\varepsilon}_{n-1}\,a+\bar{\varepsilon}_{n-1}}_{
\xi +\bar{\varepsilon}_{\mu}\,a+\bar{\varepsilon}_{n-2}}
\Psi^* (v)_\xi^{\xi+\bar{\varepsilon}_{\mu}}. 
\label{eq:psiLambda}
\end{equation}
This equation can be rewritten as follows: 
\begin{equation}
\begin{array}{cl}
&\displaystyle \Psi^* (v)_\xi^{\xi +\bar{\varepsilon}_{n-1}} 
\Lambda (u)^{\xi\,a+\bar{\varepsilon}_{n-1}}_{\xi\,a+\bar{\varepsilon}_{n-2}}-
\Lambda (u)^{\xi +\bar{\varepsilon}_{n-1}\,a+\bar{\varepsilon}_{n-1}}_{
\xi +\bar{\varepsilon}_{n-1}\,a+\bar{\varepsilon}_{n-2}}
\Psi^* (v)_\xi^{\xi +\bar{\varepsilon}_{n-1}} \\
=&\displaystyle\sum_{\mu =0}^{n-2} 
\dfrac{[u+\Delta u-v+\xi_{\mu\,n-1}]'}{[u+\Delta u-v]'}
\prod_{j\neq\mu} \dfrac{[\xi_{n-1\,j}+1]'}{[\xi_{\mu j}+1]'} 
\Lambda (v)^{\xi +\bar{\varepsilon}_{n-1}\,a+\bar{\varepsilon}_{n-1}}_{
\xi +\bar{\varepsilon}_{\mu}\,a+\bar{\varepsilon}_{n-2}}
\Psi^* (v)_\xi^{\xi+\bar{\varepsilon}_{\mu}}. 
\end{array}
\label{eq:psiLambda'}
\end{equation}
Since the tail operators on the LHS of (\ref{eq:psiLambda'}) are diagonal 
components with respect to the ground state sectors, 
the free field representation (\ref{eq:Bose-Lambda}) can be used. 
Thus, we have 
\begin{equation}
\begin{array}{cl}
& \mbox{LHS of (\ref{eq:psiLambda'})}=(-1)^{n-1}
G_K \displaystyle\oint_C \dfrac{dz'}{2\pi\sqrt{-1}z'} 
\oint_{C'} \prod_{j=1}^{n-1}\frac{dz_j}{2\pi \sqrt{-1} z_j} 
[V_{-\alpha_{n-1}}(v_{n-1}), U_{-\alpha_{n-1}}(v')] \\
\times& \displaystyle V_{-\alpha_{n-2}}(v_{n-2}) \cdots V_{-\alpha_1}(v_1) 
V_{\omega_1}(v) f(v'-u, K_{n-2\,n-1}) 
\prod_{j=0}^{n-2}f^*(v_{j}-v_{j+1},1-L_{j n-1}) G_K^{-1}, 
\end{array}
\label{eq:LHS-p*L}
\end{equation}
where $z_j=x^{2v_j}$ and $z'=x^{2v'}$. From (\ref{eq:VU-comm}) the integral 
with respect to $z_{n-1}$ of 
(\ref{eq:LHS-p*L}) can be 
evaluated by the residues at $z_{n-1}=-x^{\pm 1}z'$. 
Then the result is 
\begin{equation}
\begin{array}{cl}
&\mbox{LHS}|_{{\cal F}_{\xi +\rho, a+\bar{\varepsilon}_{n-2}+\rho}} =
\dfrac{(-1)^{n-1}}{x^{-1}-x} 
G_K \displaystyle\oint_C \dfrac{dz'}{2\pi\sqrt{-1}z'} 
\oint_{C'} \prod_{j=1}^{n-2}\frac{dz_j}{2\pi \sqrt{-1} z_j} \\ 
\times & \displaystyle\left( 
F\left( v'+\tfrac{r}{2} \right) 
W_{-\alpha_{n-1}} 
\left( v'+\tfrac{r}{2} \right) 
- F\left( v'-\tfrac{r}{2} \right) 
W_{-\alpha_{n-1}} 
\left( v'-\tfrac{r}{2} \right) \right) \\ 
\times & 
\displaystyle V_{-\alpha_{n-2}}(v_{n-2}) \cdots V_{-\alpha_1}(v_1) 
V_{\omega_1}(v) \displaystyle 
\prod_{j=0}^{n-3}f^*(v_{j}-v_{j+1},1-\xi_{j n-1}) G_K^{-1} \\
=& \dfrac{(-1)^n}{x^{-1}-x} 
G_K \displaystyle \left( \oint_{x^{-r}C} 
-\oint_{x^{r}C} \right) \dfrac{dz'}{2\pi\sqrt{-1}z'} 
\oint_{C'} \prod_{j=1}^{n-2}\frac{dz_j}{2\pi \sqrt{-1} z_j} F(v') 
W_{-\alpha_{n-1}} (v') \\ 
\times & \displaystyle V_{-\alpha_{n-2}}(v_{n-2}) \cdots V_{-\alpha_1}(v_1) 
V_{\omega_1}(v) \prod_{j=0}^{n-3}f^*(v_{j}-v_{j+1},1-\xi_{j n-1}) G_K^{-1}, 
\end{array} \label{eq:LHS-p*L2}
\end{equation}
where 
\begin{equation}
F(v')=\dfrac{[v_{n-2}-v'+\frac{r}{2}-\frac{\pi\sqrt{-1}}{2\epsilon}
-\xi_{n-2\,n-1}]'}{
[v_{n-2}-v'+\frac{r}{2}-\frac{\pi\sqrt{-1}}{2\epsilon}]'}
\dfrac{[v'-u-\frac{r+1}{2}-a_{n-2\,n-1}]}{
[v'-u-\frac{r+1}{2}]}. 
\end{equation}
The integral with respect to $z'$ of (\ref{eq:LHS-p*L2}) can be evaluated by 
the residues at $z'=-x^r z_{n-2}$ and $z'=x^{-r+1+2u}$. The former residue 
vanishes because of (\ref{eq:WV})\footnote{
When $n=2$ we use (\ref{eq:WV'}). }. 
Thus we have 
\begin{equation}
\begin{array}{rcl}
(\ref{eq:LHS-p*L2})&=&(-1)^n \displaystyle
\oint_{C'} \prod_{j=1}^{n-2}\frac{dz_j}{2\pi \sqrt{-1} z_j} 
W_{-\alpha_{n-1}} \left( u-\tfrac{r-1}{2} \right) 
V_{-\alpha_{n-2}}(v_{n-2}) \cdots V_{-\alpha_1}(v_1) 
V_{\omega_1}(v) \\
&\times&\displaystyle\prod_{j=0}^{n-2}f^*(v_{j}-v_{j+1},1-\xi_{j\,n-1})
\dfrac{[a_{n-2\,n-1}]}{(x^{-1}-x)(x^{2r}; x^{2r})_\infty^3} 
\dfrac{G_{a+\bar{\varepsilon}_{n-1}}}{G_{a+\bar{\varepsilon}_{n-2}}}. 
\end{array} \label{eq:LHS-p*L3}
\end{equation}
On (\ref{eq:LHS-p*L3}), we should read as 
$v_{n-1}=u+\frac{\pi\sqrt{-1}}{2\epsilon}$. Equating (\ref{eq:LHS-p*L3}) 
and the RHS of (\ref{eq:psiLambda'}) and 
using the identity (\ref{eq:sum=0^*}), 
we find the free filed representation of the tail operator 
\begin{equation}
\begin{array}{cl}
&\Lambda (v)^{\xi +\bar{\varepsilon}_{n-1}\,a+\bar{\varepsilon}_{n-1}}_{
\xi +\bar{\varepsilon}_{\mu}\,a+\bar{\varepsilon}_{n-2}}=
\dfrac{(-1)^{n-\mu}[a_{n-2\,n-1}]}{(x^{-1}-x)(x^{2r}; x^{2r})_\infty^3} 
\dfrac{[\xi_{\mu \, n-1}-1]'}{[1]'} G_K G'_L{}^{-1} \\ 
\times & \displaystyle
\oint_{C'} \prod_{j=\mu +1}^{n-2}\frac{dz_j}{2\pi \sqrt{-1} z_j} 
W_{-\alpha_{n-1}} \left( u-\tfrac{r-1}{2} \right) 
V_{-\alpha_{n-2}}(v_{n-2}) \cdots V_{-\alpha_{\mu +1}}(v_{\mu +1}) \\
\times & \displaystyle 
\prod_{j=\mu +1}^{n-2} f^*(v_{j}-v_{j+1},L_{\mu j})G_K^{-1}G'_L , 
\end{array} 
\label{eq:Lambda-repII}
\end{equation}
for $0\leqslant \mu\leqslant n-2$ 
with $\Delta u=-\frac{n-1}{2}+\frac{\pi\sqrt{-1}}{2\epsilon}$ and 
$v_{n-1}=u+\frac{\pi\sqrt{-1}}{2\epsilon}$. 

Let us return to eq. (\ref{eq:Lambda-psi}) with 
$\Delta u=-\frac{n-1}{2}+\frac{\pi\sqrt{-1}}{2\epsilon}$. 
By taking an appropriate linear combination of (\ref{eq:Lambda-psi}), 
we have the following relation: 
\begin{equation}
\sum_{\mu =0}^{n-1}A_\mu \Psi^* (v)^{\xi'}_{\xi' -\bar{\varepsilon}_\mu} 
\Lambda (u)^{\xi' -\bar{\varepsilon}_\mu\,a'}_{\xi\,a} =B
\Lambda (u)_{\xi +\bar{\varepsilon}_0\,a}^{\xi'\,a'} 
\Psi^* (v)^{\xi +\bar{\varepsilon}_0}_{\xi}. 
\label{eq:II-L-linearcomb}
\end{equation}
Here, the coefficients are 
$$
\begin{array}{rcl}
A_\mu &=&\displaystyle
\prod_{j=0\atop j\neq\mu}^{n-1} \dfrac{1}{[\xi'_{\mu j}]'} 
\dfrac{[u-v-\frac{n-1}{2}+\frac{\pi\sqrt{-1}}{2\epsilon}+
\bar{\xi}'_\mu -\bar{\xi}_0 +\frac{1}{n}]'}
{[u-v-\frac{n-1}{2}+\frac{\pi\sqrt{-1}}{2\epsilon}+
\bar{\xi}'_0 -\bar{\xi}_0 +\frac{1}{n}]'}
\dfrac{[\bar{\xi}'_0 -\bar{\xi}_0 +\frac{1}{n}]'}
{[\bar{\xi}'_\mu -\bar{\xi}_0 +\frac{1}{n}]'}, \\ 
B&=&\dfrac{[u-v-\frac{n-3}{2}+\frac{\pi\sqrt{-1}}{2\epsilon}]'}
{[u-v-\frac{n-3}{2}+\frac{\pi\sqrt{-1}}{2\epsilon}+
\bar{\xi}'_0 -\bar{\xi}_0 +\frac{1}{n}]'}
\displaystyle\prod_{j=1}^{n-1}\dfrac{[\xi'_{j0}]'}{
[\bar{\xi}'_j -\bar{\xi}_0 +\frac{1}{n}]'[\xi_{0j}+1]'}. 
\end{array}
$$
Consider the product 
\begin{equation}
\begin{array}{cl}
& V_{\omega_1}(v )V_{-\alpha_1}(v_1)\cdots 
V_{-\alpha_{n-2}}(v_{n-2})
W_{-\alpha_{n-2}}\left(u-\tfrac{r-1}{2}\right) \\
=& :V_{\omega_1}(v )V_{-\alpha_1}(v_1)\cdots 
V_{-\alpha_{n-2}}(v_{n-2})
W_{-\alpha_{n-2}}\left(u-\tfrac{r-1}{2}\right): \\
\times&\displaystyle\prod_{j=1}^{n-2} 
z_{j-1}^{-\frac{r}{r-1}}\dfrac{(x^{2r-1}
\frac{z_j}{z_{j-1}}; x^{2r-2})_\infty}
{(x^{-1}\frac{z_j}{z_{j-1}}; x^{2r-2})_\infty} \cdot 
z_{n-2}^{-\frac{1}{r-1}}\dfrac{(-x
\frac{x^{2u}}{z_{n-2}}; x^{2r-2})_\infty}
{(-x^{-1}\frac{x^{2u}}{z_{n-2}}; x^{2r-2})_\infty}. 
\end{array}
\label{eq:OPE-II+W}
\end{equation}
The convergence domain of (\ref{eq:OPE-II+W}) is that 
$x^{-1}|z_j |<|z_{j-1}|$ ($1\leqslant j\leqslant n-2$) and 
$|-x^{2u-1}|<|z_{n-2}|$. Thus, each term of 
the LHS of (\ref{eq:II-L-linearcomb}) has a pole at 
$z=-x^{1-n}x^{2u}$ ($v=u-\frac{n-1}{2}+\frac{\pi\sqrt{-1}}{
2\epsilon}$) because pinching occurs at the pole. 
On the other hand, the RHS of (\ref{eq:II-L-linearcomb}) 
does not have such a pole. Hence the singularities 
at $v=u-\frac{n-1}{2}+\frac{\pi\sqrt{-1}}{2\epsilon}$ on 
the RHS of (\ref{eq:II-L-linearcomb}) cancel each other: 
\begin{equation}
\sum_{\mu =0}^{n-1} 
\prod_{j=0\atop j\neq\mu}^{n-1} \dfrac{1}{[\xi'_{\mu j}]'}
\Psi^* \left(u-\tfrac{n-1}{2}+\tfrac{\pi\sqrt{-1}}{2\epsilon}\right)^{
\xi'}_{\xi' -\bar{\varepsilon}_\mu} 
\Lambda (u)^{\xi' -\bar{\varepsilon}_\mu\,a'}_{\xi\,a} =O(1). 
\label{eq:II-L-linearcomb-limit}
\end{equation}
{}From (\ref{eq:II-L-linearcomb-limit}) and (\ref{eq:sum=0^*}) 
we find the representation 
\begin{equation}
\Lambda (u)^{\xi' -\bar{\varepsilon}_\mu\,a'}_{\xi\,a}=
\oint_{C'} \prod_{j=\mu +1}^{n-1} f^* (v_j -v_{j+1}, L_{j\mu}) 
\dfrac{dz_j}{2\pi\sqrt{-1}z_j} V_{-\alpha_{\mu +1}}(v_{\mu +1})\cdots 
V_{-\alpha_{n-1}}(v_{n-1}) \cdot 
\Lambda (u)^{\xi' -\bar{\varepsilon}_{n-1}\,a'}_{\xi\,a}, 
\label{eq:Lambda-II-general}
\end{equation}
where $v_n =u+\frac{1}{2}+\frac{\pi\sqrt{-1}}{2\epsilon}$. 

In a similar way to derive 
(\ref{eq:II-L-linearcomb}) from (\ref{eq:Lambda-psi}), 
we can derive the following relation from (\ref{eq:Lambda-phi}): 
\begin{equation}
\begin{array}{cl}
&\displaystyle\sum_{\mu =0}^{n-1} 
\Lambda (u)^{\xi'\,a'}_{\xi\,a+\bar{\varepsilon}_\mu} 
\Phi (v)^{a+\bar{\varepsilon}_\mu}_a \prod_{j=0\atop j\neq\mu}^{n-1} 
\dfrac{[a_{\mu j}+1]}{[a_{\mu j}]} [u-v+\bar{a'}_{\nu}-\bar{a}_\mu +
\tfrac{1}{n}] \prod_{j=0\atop j\neq\nu}^{n-1} 
[\bar{a'}_{j}-\bar{a}_\mu +\tfrac{1}{n}] \\
=& [u-v+1] \displaystyle\prod_{j=0\atop j\neq\nu}^{n-1} 
[a'{}_{\nu\,j}] \Phi (v)^{a'}_{a'-\bar{\varepsilon}_{\nu}} 
\Lambda (u)^{\xi'\,a'-\bar{\varepsilon}_{\nu}}_{\xi\,a}. 
\end{array}
\label{eq:I-L-linearcomb}
\end{equation}
Let $v=u+1$ and take the sum over $0\leqslant\nu\leqslant n-1$. 
Then we have 
\begin{equation}
\displaystyle\sum_{\mu =0}^{n-1} A_\mu (a,a')
\Lambda (u)^{\xi'\,a'}_{\xi\,a+\bar{\varepsilon}_\mu} 
\Phi (u+1)^{a+\bar{\varepsilon}_\mu}_a 
\prod_{j=0\atop j\neq\mu}^{n-1} 
\dfrac{[(a+\bar{\varepsilon}_\mu )_{\mu j}]}{[a_{\mu j}]} =0, 
\label{eq:I-L-linearcomb-limit}
\end{equation}
where 
$$
A_\mu (a,a')=\sum_{\nu =0}^{n-1} \prod_{j=0}^{n-1} 
[(a'-\bar{\varepsilon}_{\nu})_{j}-\bar{a}_\mu ]. 
$$
{}From (\ref{eq:I-L-linearcomb-limit}) and (\ref{eq:sum=0}), 
we obtain the expression 
\begin{equation}
\begin{array}{rcl}
\Lambda (u)^{\xi'\,a'}_{\xi\,a+\bar{\varepsilon}_\mu} 
&=&
\Lambda (u)^{\xi'\,a'}_{\xi\,a+\bar{\varepsilon}_{n-1}} (-1)^{n-1-\mu} G_K 
\displaystyle\oint_C \prod_{j=\mu +1}^{n-1} \dfrac{dz_j}{2\pi\sqrt{-1}z_j} 
U_{-\alpha_{n-1}}(v_{n-1}) \cdots U_{-\alpha_{\mu +1}}(v_{\mu +1})  \\ 
&\times& 
\displaystyle\prod_{j=\mu +1}^{n-1} 
f(v_j -v_{j+1}, K_{\mu\,j}) G_K^{-1}\dfrac{A_{n-1}(a,a')}{A_{\mu}(a,a')}, 
\end{array}
\label{eq:Lambda-I-general}
\end{equation}
where $v_n =u-\frac{n-2}{2}$. 

Combining eqs. (\ref{eq:Lambda-repII}, 
\ref{eq:Lambda-II-general}, \ref{eq:Lambda-I-general}), we can 
construct a free field representations of any 
$\Lambda(u)^{\xi'\,a'}_{\xi\,a}$, in principle. 

\subsection{Form factors} 

Form factors of $(\mathbb{Z}/n\mathbb{Z})$-symmetric model are 
defined as matrix elements of some local operators. 
Consider the local operator 
\begin{equation}
{\cal O}=E^{(1)}_{\mu_1 \mu'_1} 
\cdots E^{(N)}_{\mu_N \mu'_N}, 
\label{eq:l-op0}
\end{equation}
where $E^{(j)}_{\mu_j \mu'_j}$ 
is the matrix unit on the $j$-th site. The free field representation 
of ${\cal O}$ is given by 
\begin{equation}
\hat{{\cal O}}=\Phi^{*}_{\mu_1} (u_1 )
\cdots 
\Phi^{*}_{\mu_N} (u_N )
\Phi^{\mu'_N} (u_N )
\cdots 
\Phi^{\mu'_1} (u_1 ). 
\label{eq:l-op}
\end{equation}
The corresponding form factors 
with $m$ `charged' particles are given by 
\begin{equation}
F^{(i)}_{m}({\cal O}; v_1 , \cdots , v_{m})_{
\nu_{1} \cdots \nu_{m}}=\dfrac{1}{\chi^{(i)}} \mbox{Tr}_{
{\cal H}^{(i)}}\, \left( 
\Psi^{*}_{\nu_{1}} (v_{1})
\cdots \Psi^{*}_{\nu_m} (v_m ) \hat{{\cal O}} \rho^{(i)} 
\right), 
\label{eq:d-ff}
\end{equation}
where 
\begin{equation}
\chi^{(i)}=\mbox{Tr}_{
{\cal H}^{(i)}}\, \rho^{(i)} 
=\dfrac{(x^{2n}; x^{2n})_\infty}{
(x^{2}; x^{2})_\infty}. 
\end{equation}
and $m\equiv 0$ (mod $n$). 
Note that the local operator (\ref{eq:l-op0}) commute 
with the type II vertex operators because of 
(\ref{eq:l-op}) and (\ref{eq:chiPsiPhi}). 

By using (\ref{eq:rho-rel}), (\ref{eq:T^psi}) and (\ref{eq:T^Phi}), 
we can rewrite (\ref{eq:d-ff}) as follows: 
\begin{equation}
\begin{array}{cl}
&F^{(i)}_{m}({\cal O}; v_1 , \cdots , v_{m})_{
\nu_{1} \cdots \nu_{m}} \\
=&\displaystyle\dfrac{1}{\chi^{(i)}} \sum_{\xi_1,\cdots , 
\xi_m} t'{}^*_{\nu_1} 
\left(v_1-u+\tfrac{n-1}{2}-\tfrac{\pi\sqrt{-1}}{2\epsilon}\right){}_{
\xi}^{\xi_1}\cdots t'{}^*_{\nu_m} 
\left(v_m-u+\tfrac{n-1}{2}-\tfrac{\pi\sqrt{-1}}{2\epsilon}\right){}_{
\xi_{m-1}}^{\xi_m} \\
\times&\displaystyle\displaystyle\sum_{
k\equiv l+\omega_i\atop\mbox{\scriptsize (mod $Q$)}}
\sum_{a_1\cdots a_N\atop a'_1\cdots a'_N} 
t^*_{\mu_1}(u_1-u)^{a}_{a_1} \cdots t^*_{\mu_N}(u_N-u)^{a_{N-1}}_{a_N} 
t^{\mu'_N}(u_N-u)^{a_N}_{a'_N} \cdots t^{\mu'_1}(u_1-u)^{a'_{2}}_{a'_1} \\
\times&\displaystyle \mbox{Tr}_{{\cal H}^{(i)}_{l,k}}\, 
\left( \Psi^* (v_1)^\xi_{\xi_1} \cdots \Psi^* (v_m)^{\xi_{m-1}}_{\xi_m}
\Phi^* (u_1)^{a}_{a_1} \cdots \Phi^* (u_N)^{a_{N-1}}_{a_N} 
\Phi (u_N)^{a_N}_{a'_N} \cdots \Phi (u_1)^{a'_{2}}_{a'_1}
\Lambda (u)_{\xi\,a}^{\xi_m a'_1} \dfrac{\rho^{(i)}_{l,k}}{b_l} \right), 
\end{array}
\label{eq:ff-rep}
\end{equation}
where $k=a+\rho$, $l=\xi +\rho$, and 
\begin{equation}
b_l =\left( \dfrac{
(x^{2r};x^{2r})_\infty}{(x^{2r-2};x^{2r-2})_\infty} \right)^{(n-1)(n-2)/2} 
G'_\xi . 
\label{eq:b_l}
\end{equation}
Free filed representations of the tail operators $\Lambda$'s 
have been constructed in the present paper, besides all other operators 
$\Phi$'s, $\Phi^*$'s and $\Psi^*$'s on (\ref{eq:ff-rep}) were 
given in \cite{AJMP, FKQ, Bel-corr}. 
Integral formulae can be therefore obtained for form factors 
of $(\mathbb{Z}/n\mathbb{Z})$-symmetric model, 
in principle. 

\section{Concluding remarks} 

In this paper we present vertex operator approach for form factors 
of $(\mathbb{Z}/n\mathbb{Z})$-symmetric model. For that purpose 
we constructed the free field representations of the tail operators 
$\Lambda_{\xi\,a}^{\xi' a'}$, the nonlocal operators which relate 
the physical quantities of $(\mathbb{Z}/n\mathbb{Z})$-symmetric model 
and $A^{(1)}_{n-1}$-model. As a result, we can obtain 
the integral formulae for form factors 
of $(\mathbb{Z}/n\mathbb{Z})$-symmetric model, in principle. 

Our approach is based on some assumptions. We assumed that the vertex 
operator algebra (\ref{eq:rho-rel}--\ref{eq:T_phi}) and 
(\ref{eq:T^psi}--\ref{eq:T_Psi}) correctly describes 
the intertwining relation between $(\mathbb{Z}/n\mathbb{Z})$-symmetric model 
and $A^{(1)}_{n-1}$-model. We also assumed that 
the free field representations (\ref{eq:Lambda-repII}, 
\ref{eq:Lambda-II-general}, \ref{eq:Lambda-I-general}) provide 
relevant representations of the vertex operator algebra. 
As a consistency check of our bosonization scheme, 
it is thus important to derive closed expressions for form factors of 
some simple local operators by performing the integrals on (\ref{eq:ff-rep}). 
We wish to address the problem 
in a separate paper. 

Before ending the present paper, we should add one thing. 
In order to find the free field representations of the tail operators 
(\ref{eq:Lambda-repII}), we used the correct commutation relation 
(\ref{eq:VU-comm}). In our previous paper \cite{FKQ} 
we proved (\ref{eq:Wchipsiphi}) by using the commutativity of 
$U_{-\alpha_{j}}(v)$ and $V_{-\alpha_{j}}(v')$, instead of 
(\ref{eq:VU-comm}). In Appendix C we thus prove (\ref{eq:Wchipsiphi}) 
on the basis of (\ref{eq:VU-comm}). 

\section*{Acknowledgements} 

We would like to thank R. Inoue, H. Konno 
and Y. Takeyama for discussion and their interests in the present work.

\appendix 

\section{Appendix A ~~ Definitions of the models concerned}

\subsection{Belavin's vertex model}

In the original papers \cite{Bela,RT}, 
the $R$-matrix in the disordered phase is given. 
For the present purpose, we need the following $R$-matrix: 
\begin{equation}
\begin{array}{rcl}
R(v)&=&\dfrac{[1]}{[1-v]}r_1 (v)\overline{R}(v), \\
\overline{R}(v)^{ik}_{jl}&=&
\displaystyle\frac{h(v)
\vartheta \left[\begin{array}{c} \frac{1}{2} \\ 
\frac{1}{2}+\frac{k-i}{n} \end{array} \right]
\left( \dfrac{1-v}{nr} ; 
\dfrac{\pi\sqrt{-1}}{n\epsilon r} \right)}
{\vartheta \left[\begin{array}{c} \frac{1}{2} \\ 
\frac{1}{2}+\frac{j-k}{n} \end{array} \right]
\left( \dfrac{v}{nr} ; 
\dfrac{\pi\sqrt{-1}}{n\epsilon r} \right)
\vartheta \left[\begin{array}{c} \frac{1}{2} \\ 
\frac{1}{2}+\frac{j-i}{n} \end{array} \right]
\left( \dfrac{1}{nr} ; 
\dfrac{\pi\sqrt{-1}}{n\epsilon r} \right)}\delta^{i+k}_{j+l\;(\mbox{mod}\,n)}, 
\end{array}
\label{eq:Bel}
\end{equation}
where $r_1 (v)$ is defined by (\ref{eq:g-def}), and 
$$
h(v)=\prod_{j=0}^{n-1} 
\vartheta \left[\begin{array}{c} \frac{1}{2} \\ 
\frac{1}{2}+\frac{j}{n} \end{array} \right]
\left( \frac{v}{nr} ; 
\frac{\pi\sqrt{-1}}{n\epsilon r} \right)\left/
\;\prod_{j=1}^{n-1} \vartheta \left[\begin{array}{c} \frac{1}{2} \\ 
\frac{1}{2}+\frac{j}{n} \end{array} \right]\right.
\left( 0 ; 
\frac{\pi\sqrt{-1}}{n\epsilon r} \right). 
$$
We assume that the parameters $v$, $\epsilon$ and $r$ lie 
in the so-called principal regime: 
\begin{equation}
\epsilon >0, ~~ r>n-1, ~~ 0<v<1. 
\label{eq:principal}
\end{equation}
Note that the weights (\ref{eq:Bel}) reproduce those of the eight-vertex 
model in the principal regime when $n=2$ \cite{ESM}.

\subsection{The weight lattice and the root lattice 
of $A^{(1)}_{n-1}$}

Let $V=\mathbb{C}^n$ and 
$\{ \varepsilon _\mu \}_{0 \leqslant \mu \leqslant n-1}$ be 
the standard orthonormal basis as before. 
The weight lattice of $A^{(1)}_{n-1}$ 
is defined as follows: 
\begin{equation}
P=\bigoplus_{\mu =0}^{n-1} 
\mathbb{Z} \bar{\varepsilon}_\mu , 
\label{eq:wt-lattice}
\end{equation}
where 
$$
\bar{\varepsilon}_\mu =
\varepsilon _\mu -\varepsilon , 
~~~~\varepsilon =\frac{1}{n}\sum_{\mu =0}^{n-1} 
\varepsilon _\mu . 
$$
We denote the fundamental weights 
by $\omega_\mu\,(1\leqslant \mu \leqslant n-1)$ 
$$
\omega_\mu =\sum_{\nu =0}^{\mu -1}
\bar{\varepsilon }_\nu , 
$$
and also denote the simple roots by 
$\alpha_\mu \,(1\leqslant \mu \leqslant n-1)$ 
$$
\alpha_\mu =\varepsilon_{\mu-1} -
\varepsilon _{\mu}=\bar{\varepsilon}_{\mu -1}
-\bar{\varepsilon}_{\mu}. 
$$
The root lattice of $A^{(1)}_{n-1}$ 
is defined as follows: 
\begin{equation}
Q=\bigoplus_{\mu =1}^{n-1} 
\mathbb{Z} \alpha_\mu , 
\label{eq:rt-lattice}
\end{equation}

For $a\in P$ we set 
\begin{equation}
a_{\mu\nu}=\bar{a}_\mu-\bar{a}_\nu , ~~~~ 
\bar{a}_\mu =\langle a+\rho , 
\varepsilon_\mu \rangle =\langle a+\rho , 
\bar{\varepsilon}_\mu \rangle , ~~~~ 
\rho =\sum_{\mu =1}^{n-1} \omega_\mu . 
\end{equation}

In this paper we admit not only the case $a\in P$ 
but also the case $a\in {\mathfrak h}^*:=
\displaystyle \mathbb{C}\omega_0 \oplus \mathbb{C} 
\omega_1 \oplus \cdots \oplus \mathbb{C} 
\omega_{n-1}$. For $r>n-1$, 
let $\displaystyle\sum_{\mu=0}^{n-1} k^\mu =r$, where 
$a+\rho =\displaystyle\sum_{\mu=0}^{n-1} k^\mu \omega_\mu$, 
then we denote $a\in {\mathfrak h}^*_{r-n}$. 

\subsection{The $A^{(1)}_{n-1}$ face model}

An ordered pair $(a,b) \in {\mathfrak h}^*{}^2_{r-n}$ 
is called {\it admissible} if $b=a+\bar{\varepsilon}_\mu$, 
for a certain $\mu\,(0\leqslant \mu \leqslant n-1)$. 
Non-zero Boltzmann weights are parametrized in terms of 
the elliptic theta function of the spectral parameter $v$ 
as follows: 
\begin{eqnarray}
W
\left[ \left. \begin{array}{cc} 
a + 2 \bar{\varepsilon }_\mu & a+\bar{\varepsilon }_\mu \\ 
a+\bar{\varepsilon }_\mu & a \end{array} \right| 
v  \right] 
& = & r_1 (v), \label{eq:BW1} \\[3mm]
W
\left[ \left. \begin{array}{cc} 
a+\bar{\varepsilon }_\mu +\bar{\varepsilon }_\nu 
& a+\bar{\varepsilon }_\mu \\ 
a+\bar{\varepsilon }_\nu & a 
\end{array} \right| v \right] & = & r_1 (v)
\dfrac{[v][a_{\mu\nu}+1]}{[1-v][a_{\mu\nu}]} 
~~~~(\mu \neq \nu ), \label{eq:BW2} \\[3mm]
W
\left[ \left. \begin{array}{cc} 
a+\bar{\varepsilon }_\mu +\bar{\varepsilon }_\nu 
& a+\bar{\varepsilon }_\mu \\ 
a+\bar{\varepsilon }_\mu & a 
\end{array} \right| v \right] & = & r_1 (v)
\dfrac{[1][v+a_{\mu\nu}]}{[1-v][a_{\mu\nu}]} 
~~~~(\mu\neq \nu), \label{eq:BW3} 
\end{eqnarray}
where $r_1 (v)$ is defined by (\ref{eq:g-def}). In this paper 
we consider so-called Regime III in the model, i.e., 
$0<v<1$.

\section{Appendix B ~~ OPE formulae and commutation relations}

In this Appendix we list some useful formulae for 
the basic bosons. In what follows we denote 
$z=x^{2v}$, $z'=x^{2v'}$. 

First, useful OPE formulae are: 
\begin{eqnarray}
U_{\omega_1}(v)U_{\omega_j}(v')&=& z^{\frac{r-1}{r}\frac{n-j}{n}} g_j(z'/z)
:U_{\omega_1}(v)U_{\omega_j}(v'):, \\
U_{\omega_j}(v)U_{\omega_1}(v')&=& z^{\frac{r-1}{r}\frac{n-j}{n}} g_j(z'/z)
:U_{\omega_j}(v)U_{\omega_1}(v'):, \\
U_{\omega_j}(v)U_{-\alpha_j}(v')&=&z^{-\frac{r-1}{r}} 
\dfrac{(x^{2r-1}z'/z; x^{2r})_\infty}{(xz'/z; x^{2r})_\infty}
:U_{\omega_j}(v)U_{-\alpha_j}(v'):, \label{eq:OjAj-prod} \\
U_{-\alpha_j}(v)U_{\omega_j}(v')&=&z^{-\frac{r-1}{r}} 
\dfrac{(x^{2r-1}z'/z; x^{2r})_\infty}{(xz'/z; x^{2r})_\infty}
:U_{-\alpha_j}(v)U_{\omega_j}(v'):, \\
U_{-\alpha_{j}}(v)U_{-\alpha_{j\pm 1}}(v')&=&z^{-\frac{r-1}{r}} 
\dfrac{(x^{2r-1}z'/z; x^{2r})_\infty}{(xz'/z; x^{2r})_\infty}
:U_{-\alpha_{j}}(v)U_{-\alpha_{j\pm 1}}(v'):, \label{eq:AjAj+1-prod} \\
U_{-\alpha_{j}}(v)U_{-\alpha_{j}}(v')&=&z^{\frac{2(r-1)}{r}} \left( 
1-\dfrac{z'}{z}\right) 
\dfrac{(x^{2}z'/z; x^{2r})_\infty}{(x^{2r-2}z'/z; x^{2r})_\infty}
:U_{-\alpha_{j}}(v)U_{-\alpha_{j}}(v'):, 
\end{eqnarray}
\begin{eqnarray}
V_{\omega_1}(v)V_{\omega_j}(v')&=& z^{\frac{r}{r-1}\frac{n-j}{n}} g^*_j(z'/z)
:V_{\omega_1}(v)V_{\omega_j}(v'):, \\
V_{\omega_j}(v)V_{\omega_1}(v')&=& z^{\frac{r}{r-1}\frac{n-j}{n}} g^*_j(z'/z)
:V_{\omega_j}(v)V_{\omega_1}(v'):, \\
V_{\omega_j}(v)V_{-\alpha_j}(v')&=&z^{-\frac{r}{r-1}} 
\dfrac{(x^{2r-1}z'/z; x^{2r-2})_\infty}{(x^{-1}z'/z; x^{2r-2})_\infty}
:V_{\omega_j}(v)V_{-\alpha_j}(v'):, \label{eq:OjAj-prod'} \\
V_{-\alpha_j}(v)V_{\omega_j}(v')&=&z^{-\frac{r}{r-1}} 
\dfrac{(x^{2r-1}z'/z; x^{2r-2})_\infty}{(x^{-1}z'/z; x^{2r-2})_\infty}
:V_{-\alpha_j}(v)V_{\omega_j}(v'):, \\
V_{-\alpha_{j}}(v)V_{-\alpha_{j\pm 1}}(v')&=&z^{-\frac{r}{r-1}} 
\dfrac{(x^{2r-1}z'/z; x^{2r-2})_\infty}{(x^{-1}z'/z; x^{2r-2})_\infty}
:V_{-\alpha_{j}}(v)V_{-\alpha_{j\pm 1}}(v'):, \label{eq:AjAj+1-prod'} \\
V_{-\alpha_{j}}(v)V_{-\alpha_{j}}(v')&=&z^{\frac{2r}{r-1}} \left( 
1-\dfrac{z'}{z}\right) 
\dfrac{(x^{-2}z'/z; x^{2r-2})_\infty}{(x^{2r}z'/z; x^{2r-2})_\infty}
:V_{-\alpha_{j}}(v)V_{-\alpha_{j}}(v'):, 
\end{eqnarray}
\begin{eqnarray}
V_{\omega_j}(v)U_{\omega_j}(v')&=& z^{-\frac{j(n-j)}{n}} \rho_j (z'/z) 
:V_{\omega_1}(v)U_{\omega_j}(v'):, \\
U_{\omega_j}(v)V_{\omega_j}(v')&=& z^{-\frac{j(n-j)}{n}} \rho_j (z'/z) 
:U_{\omega_j}(v)V_{\omega_j}(v'):, \\
V_{\omega_{j}}(v)U_{-\alpha_{j}}(v')&=&z\left( 1+\dfrac{z'}{z} \right) 
:V_{\omega_{j}}(v)U_{-\alpha_{j}}(v'): \,=\, 
U_{-\alpha_{j}}(v')V_{\omega_{j}}(v), \label{eq:VoUa} \\
U_{\omega_{j}}(v)V_{-\alpha_{j}}(v')&=&z\left( 1+\dfrac{z'}{z} \right) 
:U_{\omega_{j}}(v)V_{-\alpha_{j}}(v'): \,=\, 
V_{-\alpha_{j}}(v')U_{\omega_{j}}(v), \\
V_{-\alpha_{j}}(v)U_{-\alpha_{j\pm 1}}(v')&=&z\left( 1+\dfrac{z'}{z} \right) 
:V_{-\alpha_{j}}(v)U_{-\alpha_{j\pm 1}}(v'): \,=\, 
U_{-\alpha_{j\pm 1}}(v')V_{-\alpha_{j}}(v), \label{eq:VaUa} \\
V_{-\alpha_{j}}(v)U_{-\alpha_{j}}(v')&=&
\dfrac{:V_{-\alpha_{j}}(v)U_{-\alpha_{j}}(v'):}{
z^2(1+\frac{xz'}{z})(1+\frac{x^{-1}z'}{z})}, \label{eq:VU-normal} \\[3mm]
U_{-\alpha_{j}}(v)V_{-\alpha_{j}}(v')&=&
\dfrac{:U_{-\alpha_{j}}(v)V_{-\alpha_{j}}(v'):}{
z{}^2(1+\frac{xz'}{z})(1+\frac{x^{-1}z'}{z})}, \label{eq:UV-normal} 
\end{eqnarray}
where $g_j (z)$, $g^*_j (z)$ and $\rho_j (z)$ 
are defined by (\ref{eq:g-def}), (\ref{eq:g*-def}) and 
(\ref{eq:chi-def}). From these, we obtain the following 
commutation relations: 
\begin{eqnarray}
U_{\omega_1}(v)U_{\omega_j}(v')&=&r_j (v-v')
U_{\omega_j}(v')U_{\omega_1}(v), \\
U_{-\alpha_j}(v)U_{\omega_j}(v')&=&-f(v-v',0)
U_{\omega_j}(v')U_{-\alpha_j}(v), \\
U_{-\alpha_j}(v)U_{-\alpha_{j\pm 1}}(v')&=&-f(v-v',0)
U_{-\alpha_{j\pm 1}}(v')U_{-\alpha_j}(v), \\
U_{-\alpha_j}(v)U_{-\alpha_j}(v')&=&h(v-v')
U_{-\alpha_j}(v')U_{-\alpha_j}(v), 
\end{eqnarray}
\begin{eqnarray}
V_{\omega_1}(v)V_{\omega_j}(v')&=&r^*_j (v-v')
V_{\omega_j}(v')V_{\omega_1}(v), \\
V_{-\alpha_j}(v)V_{\omega_j}(v')&=&-f^*(v-v',0)
V_{\omega_j}(v')V_{-\alpha_j}(v), \\
V_{-\alpha_j}(v)V_{-\alpha_{j\pm 1}}(v')&=&-f^*(v-v',0)
V_{-\alpha_{j\pm 1}}(v')V_{-\alpha_j}(v), \\
V_{-\alpha_j}(v)V_{-\alpha_j}(v')&=&h^*(v-v')
V_{-\alpha_j}(v')V_{-\alpha_j}(v), 
\end{eqnarray}
\begin{eqnarray}
U_{\omega_{j}}(v)V_{\omega_{j}}(v')&=&\chi_j (v-v') 
V_{\omega_{j}}(v')U_{\omega_{j}}(v), \label{eq:chiVU} \\
\mbox{[}V_{-\alpha_{j}}(v), U_{-\alpha_{j}}(v')\mbox{]}&=&
\dfrac{\delta (\frac{z}{-xz'})-\delta (\frac{z'}{-xz})}{
(x^{-1}-x)zz'}
:V_{-\alpha_{j}}(v)U_{-\alpha_{j}}(v'):, 
\label{eq:VU-comm}
\end{eqnarray}
where $r_j(v)$, $r^*_j(v)$, $\chi_j(v)$, $f(v,w)$, 
$h(v)$, $f^*(v,w)$ and $h^*(v)$ are defined by 
(\ref{eq:g-def}), (\ref{eq:g*-def}), 
(\ref{eq:chi-def}), (\ref{eq:fh-def}) and (\ref{eq:f*h*-def}), and 
the $\delta$-function is defined by the following formal 
power series 
$$
\delta (z)=\sum_{n\in \mathbb{Z}} z^n. 
$$
The commutation relation (\ref{eq:VU-comm}) can be derived from 
(\ref{eq:VU-normal}), (\ref{eq:UV-normal}) 
and the identity 
$$
\dfrac{1}{z^2
(1+\frac{xz'}{z})(1+\frac{x^{-1}z'}{z})}-
\dfrac{1}{z'{}^2
(1+\frac{xz}{z'})(1+\frac{x^{-1}z}{z'})}=
\dfrac{\delta (\frac{z}{-xz'})-\delta (\frac{z'}{-xz})}{
(x^{-1}-x)zz'}. 
$$
The relation (\ref{eq:VU-comm}) can be 
practically understood 
as follows. 
Let us compare the integrals 
\begin{equation}
\oint \dfrac{dz}{2\pi\sqrt{-1}} V_{-\alpha_j}(v)U_{-\alpha_j}(v') F(v,v'), 
\label{eq:VUf}
\end{equation}
and 
\begin{equation}
\oint \dfrac{dz}{2\pi\sqrt{-1}} U_{-\alpha_j}(v')V_{-\alpha_j}(v) F(v,v'), 
\label{eq:UVf}
\end{equation}
where $F(u,v)$ is an appropriate function. 
Note that the normal order product expansion (\ref{eq:VU-normal}) 
is valid for $|z|>|-x^{\pm 1}z'|$ while 
(\ref{eq:UV-normal}) is valid for $|z'|>|-x^{\pm 1}z|$. 
Thus, the integral contour of 
(\ref{eq:VUf}) encircles the poles $-x^{\pm 1}z'$, but 
that of (\ref{eq:UVf}) does not encircle them. 
The difference between (\ref{eq:VUf}) and (\ref{eq:UVf}) can be 
therefore evaluated by the 
residues at $z=-x^{\pm 1}z'$. 

Finally, we list the OPE formulae for $W_{-\alpha_j}(v)$ and 
other basic operators: 
\begin{eqnarray}
W_{-\alpha_{j}}(v)V_{-\alpha_{j\pm 1}}(v')&=& 
-(-z)^{-\frac{1}{r-1}} \dfrac{(-x^{r}z'/z; x^{2r-2})_\infty}
{(-x^{r-2}z'/z; x^{2r-2})_\infty}
:W_{-\alpha_{j}}(v)V_{-\alpha_{j\pm 1}}(v'):, \label{eq:WV-formula1} \\
V_{-\alpha_{j\pm 1}}(v)W_{-\alpha_{j}}(v')&=& 
z^{-\frac{1}{r-1}} \dfrac{(-x^{r}z'/z; x^{2r-2})_\infty}
{(-x^{r-2}z'/z; x^{2r-2})_\infty}
:V_{-\alpha_{j\pm 1}}(v)W_{-\alpha_{j}}(v'):, \label{eq:VWformula1} \\
V_{\omega_j}(v)W_{-\alpha_{j}}(v')&=& 
z^{-\frac{1}{r-1}} \dfrac{(-x^{r}z'/z; x^{2r-2})_\infty}
{(-x^{r-2}z'/z; x^{2r-2})_\infty}
:V_{\omega_j}(v)W_{-\alpha_{j}}(v'):, \label{eq:VWformula2} \\
W_{-\alpha_{j}}(v)V_{\omega_j}(v')&=& 
-(-z)^{-\frac{1}{r-1}} \dfrac{(-x^{r}z'/z; x^{2r-2})_\infty}
{(-x^{r-2}z'/z; x^{2r-2})_\infty}
:W_{-\alpha_{j}}(v)V_{\omega_j}(v'):, \label{eq:WV-formula2} 
\end{eqnarray}
\begin{eqnarray}
U_{-\alpha_{j\pm 1}}(v)W_{-\alpha_{j}}(v')&=& 
z^{\frac{1}{r}} \dfrac{(x^{r-1}z'/z; x^{2r})_\infty}
{(x^{r+1}z'/z; x^{2r})_\infty}
:U_{-\alpha_{j\pm 1}}(v)W_{-\alpha_{j}}(v'):, \label{eq:UW-formula1} \\
W_{-\alpha_{j}}(v)U_{-\alpha_{j\pm 1}}(v')&=& 
-z^{\frac{1}{r}} \dfrac{(x^{r-1}z'/z; x^{2r})_\infty}
{(x^{r+1}z'/z; x^{2r})_\infty}
:W_{-\alpha_{j}}(v)U_{-\alpha_{j\pm 1}}(v'):, \label{eq:WU-formula1} \\
U_{\omega_j}(v)W_{-\alpha_{j}}(v')&=& 
z^{\frac{1}{r}} \dfrac{(x^{r-1}z'/z; x^{2r})_\infty}
{(x^{r+1}z'/z; x^{2r})_\infty}
:U_{\omega_j}(v)W_{-\alpha_{j}}(v'):, \label{eq:UW-formula2} \\
W_{-\alpha_{j}}(v)U_{\omega_j}(v')&=& 
-z^{\frac{1}{r}} \dfrac{(x^{r-1}z'/z; x^{2r})_\infty}
{(x^{r+1}z'/z; x^{2r})_\infty}
:W_{-\alpha_{j}}(v)U_{\omega_j}(v'):. \label{eq:WU-formula2}
\end{eqnarray}
From these, we obtain 
\begin{eqnarray}
W_{-\alpha_j}\left( v+\tfrac{r}{2}-\tfrac{\pi\sqrt{-1}}{2\epsilon} 
\right) V_{-\alpha_{j\pm 1}}(v)&=~0~=&
V_{-\alpha_{j\pm 1}}(v)
W_{-\alpha_j}\left( v-\tfrac{r}{2}-\tfrac{\pi\sqrt{-1}}{2\epsilon} 
\right) , \label{eq:WV} \\
W_{-\alpha_j}\left( v+\tfrac{r}{2}-\tfrac{\pi\sqrt{-1}}{2\epsilon} 
\right) V_{\omega_j}(v)
&=~0~=&V_{\omega_j}(v)
W_{-\alpha_j}\left( v-\tfrac{r}{2}-\tfrac{\pi\sqrt{-1}}{2\epsilon} 
\right), \label{eq:WV'} \\
U_{-\alpha_{j\pm 1}}(v)
W_{-\alpha_j}\left( v-\tfrac{r-1}{2}\right) &=~0~=&
W_{-\alpha_j}\left( v+\tfrac{r-1}{2}\right)U_{-\alpha_{j\pm 1}}(v), 
\label{eq:UW} \\
U_{\omega_j}(v)W_{-\alpha_j}\left( v-\tfrac{r-1}{2}\right)&=~0~=&
W_{-\alpha_j}\left( v+\tfrac{r-1}{2}\right)U_{\omega_j}(v). 
\label{eq:UW'} 
\end{eqnarray}

\section{Appendix C ~~ Commutation relations of $\Phi (u)_a^{a'}$ 
and $\Psi^* (v)_\xi^{\xi'}$} 

In this appendix, we give a remark on the commutation relation 
(\ref{eq:Wchipsiphi}). In \cite{FKQ} we proved (\ref{eq:Wchipsiphi}) 
on the assumption of the commutativity of 
$U_{-\alpha_{j}}(v)$ and $V_{-\alpha_{j}}(v')$. From 
(\ref{eq:VU-comm}), however, 
$U_{-\alpha_{j}}(v)$ and $V_{-\alpha_{j}}(v')$ commute at all points 
but at $v'=v\pm \frac{1}{2}+\frac{\pi\sqrt{-1}}{2\epsilon}$. Nevertheless, 
(\ref{eq:Wchipsiphi}) holds, which we will briefly 
show in this appendix. 

Let $a'-a=\bar{\varepsilon}_\mu$ and $\xi'-\xi =\bar{\varepsilon}_\nu$ 
on (\ref{eq:Wchipsiphi}). We assume that $\mu \leqslant \nu$. 
(The case $\mu >\nu$ can be similarly proved.) 
When $\mu =0$, (\ref{eq:Wchipsiphi}) 
follows from (\ref{eq:VoUa}--\ref{eq:VaUa}) and (\ref{eq:chiVU}). 
When $\mu =1$, the difference of the both sides of (\ref{eq:Wchipsiphi}) 
can be calculated as follows: 
\begin{equation}
\begin{array}{cl}
&\Phi (u)^{a+\bar{\varepsilon}_1}_a 
\Psi^* (v)^{\xi +\bar{\varepsilon}_\nu}_{\xi}-
\chi (u-v)\Psi^* (v)^{\xi +\bar{\varepsilon}_\nu}_{\xi}
\Phi (v)^{a+\bar{\varepsilon}_1}_a \\
=& U_{\omega_1}(u)V_{\omega_1} (v) 
\displaystyle\oint_{C} \dfrac{dz'_1}{2\pi\sqrt{-1}z'_1} 
\oint_{C'} \prod_{j=1}^\nu \dfrac{dz_j}{2\pi\sqrt{-1}z_j} 
[U_{-\alpha_1} (u_1), V_{-\alpha_1} (v_1)] \\
\times& V_{-\alpha_2} (v_2) \cdots V_{-\alpha_\nu} (v_\nu ) 
f(u_1-u, K_{01}) \displaystyle\prod_{j=0\atop j\neq 1}^{n-1} 
[K_{j1}]^{-1} \prod_{j=0}^{\nu -1} 
f^* (v_{j+1}-v_j, L_{j\nu}), 
\end{array} \label{eq:dif-Wchipsiphi}
\end{equation}
where $z_j =x^{2v_j}$ and $z'_j =x^{2u_j}$. From (\ref{eq:VU-comm}) 
the integral with respect to $z_1$ of (\ref{eq:dif-Wchipsiphi}) 
can be evaluated by the residues at $z_1=-x^{\pm 1}z'_1$. 
Repeating similar calculations 
performed in section 4.4, the RHS of (\ref{eq:dif-Wchipsiphi}) 
can be rewritten as a total difference of such a form 
\begin{equation}
U_{\omega_1}(u)V_{\omega_1} (v) 
\displaystyle\left( \oint_{x^{-r}C}-\oint_{x^rC} \right) 
\dfrac{dz'_1}{2\pi\sqrt{-1}z'_1} 
\oint_{C'} \prod_{j=2}^\nu \dfrac{dz_j}{2\pi\sqrt{-1}z_j} 
W_{-\alpha_1} (u_1 ) 
V_{-\alpha_2} (v_2) \cdots V_{-\alpha_\nu} (v_\nu ) G(u_1 ), 
\label{eq:dif-Wchipsiphi-result}
\end{equation}
where 
$$
G\left( u_1+\tfrac{r}{2}\right) =\dfrac{1}{x^{-1}-x} f(u_1-u, a_{01}) 
\left. \displaystyle\prod_{j=0\atop j\neq 1}^{n-1} 
[a_{j1}]^{-1} \prod_{j=0}^{\nu -1} 
f^* (v_{j+1}-v_j, \xi_{j\nu})\right|_{v_1 =u_1+\frac{1}{2}+
\frac{\pi\sqrt{-1}}{2\epsilon}}. 
$$
In the present case, there are at most three poles at 
$u_1 =u-\frac{r-1}{2}, v-\frac{r}{2}-\frac{\pi\sqrt{-1}}{2\epsilon}, 
v_2 +\frac{r}{2}-\frac{\pi\sqrt{-1}}{2\epsilon}$, 
inside the contour for $z'_1$-integration. The residues at 
those three points vanish because of (\ref{eq:UW'}), 
(\ref{eq:WV'}), and (\ref{eq:WV}), respectively. Therefore we have 
$$
\Phi (u)^{a+\bar{\varepsilon}_1}_a 
\Psi^* (v)^{\xi +\bar{\varepsilon}_\nu}_{\xi}-
\chi (u-v)\Psi^* (v)^{\xi +\bar{\varepsilon}_\nu}_{\xi}
\Phi (v)^{a+\bar{\varepsilon}_1}_a =0. 
$$
When $\mu \geqslant 2$, 
the difference of the both sides of (\ref{eq:Wchipsiphi}) 
can be calculated as follows: 
\begin{equation}
\Phi (u)^{a+\bar{\varepsilon}_\mu}_a 
\Psi^* (v)^{\xi +\bar{\varepsilon}_\nu}_{\xi}-
\chi (u-v)\Psi^* (v)^{\xi +\bar{\varepsilon}_\nu}_{\xi}
\Phi (v)^{a+\bar{\varepsilon}_\mu}_a \\
= \sum_{\lambda =1}^\mu 
\oint_{C} \prod_{j=1}^\mu \dfrac{dz'_j}{2\pi\sqrt{-1}z'_j} 
\oint_{C'} \prod_{j'=1}^\nu \dfrac{dz_{j'}}{2\pi\sqrt{-1}z_{j'}} X_\lambda , 
\label{eq:dif-Wchipsiphi2}
\end{equation}
where 
\begin{equation}
\begin{array}{rcl}
X_\lambda &=& U_{\omega_1}(u)V_{\omega_1} (v) 
\displaystyle V_{-\alpha_1} (v_1)U_{-\alpha_1} (v_1) \cdots 
V_{-\alpha_{\lambda -1}} (v_{\lambda -1}) 
U_{-\alpha_{\lambda -1}} (u_{\lambda-1}) 
[U_{-\alpha_\lambda} (u_\lambda ), V_{-\alpha_\lambda} (v_\lambda )] \\
&\times& U_{-\alpha_{\lambda +1}} (u_{\lambda +1})
V_{-\alpha_{\lambda +1}} (v_{\lambda +1}) 
\cdots \displaystyle U_{-\alpha_{\mu}} (u_{\mu})V_{-\alpha_{\mu}} (v_{\mu}) 
\cdots V_{-\alpha_\nu} (v_\nu ) \\
&\times&\displaystyle \prod_{j=0}^{\mu -1} f(u_{j+1}-u_j, K_{j\mu}) 
\displaystyle\prod_{j=0\atop j\neq \mu}^{n-1} 
[K_{j\mu}]^{-1} \prod_{j'=0}^{\nu -1} 
f^* (v_{j'+1}-v_{j'}, L_{j'\nu}). 
\end{array} \label{eq:dif-Wchipsiphi2-detail}
\end{equation}
{}From (\ref{eq:VU-comm}) 
the integral with respect to $z_\lambda$ of $X_\lambda$ 
can be evaluated by the residues at $z_\lambda =-x^{\pm 1}z'_\lambda$. 
Similarly to (\ref{eq:dif-Wchipsiphi-result}), 
the result 
can be rewritten as a total difference of such a form 
\begin{equation}
\begin{array}{cl}
&\displaystyle\oint_{C} \prod_{j=1}^\mu \dfrac{dz'_j}{2\pi\sqrt{-1}z'_j} 
\oint_{C'} \prod_{j'=1}^\nu \dfrac{dz_{j'}}{2\pi\sqrt{-1}z_{j'}} X_\lambda 
=U_{\omega_1}(u)V_{\omega_1} (v) 
\displaystyle\left( \oint_{x^{-r}C}-\oint_{x^rC} \right) 
\dfrac{dz'_\lambda}{2\pi\sqrt{-1}z'_\lambda} \\
\times&\displaystyle\oint_{C} 
\prod_{j=1\atop j\neq\lambda}^\mu \dfrac{dz'_j}{2\pi\sqrt{-1}z'_j} 
\oint_{C'} \prod_{j'=1\atop j'\neq\lambda}^\nu 
\dfrac{dz_{j'}}{2\pi\sqrt{-1}z_{j'}}
V_{-\alpha_1} (v_1)U_{-\alpha_1} (v_1) \cdots 
V_{-\alpha_{\lambda -1}} (v_{\lambda -1}) 
U_{-\alpha_{\lambda -1}} (u_{\lambda-1}) \\
\times&\displaystyle W_{-\alpha_\lambda} (u_\lambda ) 
U_{-\alpha_{\lambda +1}} (u_{\lambda +1})
V_{-\alpha_{\lambda +1}} (v_{\lambda +1}) 
\cdots \displaystyle U_{-\alpha_{\mu}} (u_{\mu})V_{-\alpha_{\mu}} (v_{\mu}) 
\cdots V_{-\alpha_\nu} (v_\nu ) G_\lambda (u_\lambda ), 
\end{array}
\end{equation}
where 
$$
G_\lambda \left( u_\lambda +\tfrac{r}{2}\right) =\dfrac{1}{x^{-1}-x} 
\prod_{j=0}^{\mu -1} f(u_{j+1}-u_j, a_{j\mu}) 
\left. \displaystyle\prod_{j=0\atop j\neq \mu}^{n-1} 
[a_{j\mu}]^{-1} \prod_{j'=0}^{\nu -1} 
f^* (v_{j'+1}-v_{j'}, \xi_{j'\nu})\right|_{v_\lambda =u_\lambda 
+\frac{1}{2}+\frac{\pi\sqrt{-1}}{2\epsilon}}. 
$$
In the present case, there are at most four poles at 
$u_\lambda =u_{\lambda\pm 1}\pm\frac{r-1}{2}, 
v_{\lambda\pm 1}\pm\frac{r}{2}-\frac{\pi\sqrt{-1}}{2\epsilon}$, 
inside the contour for $z'_\lambda$-integration. The residues at 
those four points vanish because of (\ref{eq:UW}) 
and (\ref{eq:WV}), respectively. (When $\lambda =1$, we also use 
(\ref{eq:UW'}) and (\ref{eq:WV'}) as well as (\ref{eq:UW}) 
and (\ref{eq:WV}).) Therefore we prove (\ref{eq:Wchipsiphi}) 
for $\mu\geqslant 2$.

\end{document}